\documentclass[prb,twocolumn,aps,superscriptaddress,floatfix]{revtex4-1}
\usepackage{amsmath}
\usepackage{amssymb}
\usepackage{amsfonts}
\usepackage{bm}
\usepackage{graphicx}
\usepackage{color}
\usepackage{ulem}
\usepackage[svgnames]{xcolor}
\usepackage[colorlinks=true,citecolor=blue,linkcolor=blue,urlcolor=blue]{hyperref}

\begin{document}
%\linenumbers % Line numbering: for internal purposes

\title{Amplitude dynamics of charge density wave in
LaTe$_3$:
theoretical description of pump-probe experiments
}
\author{Pavel E. Dolgirev} 
\email{p\_dolgirev@g.harvard.edu}
\affiliation{Skolkovo Institute of Science and Technology, Skolkovo
Innovation Center, 3 Nobel St., Moscow 143026, Russia}
\affiliation{Department of Physics, Harvard University, Cambridge,
Massachusetts, 02138, USA}

\author{A.V. Rozhkov}
\email{arozhkov@gmail.com}
\affiliation{Skolkovo Institute of Science and Technology, Skolkovo
Innovation Center, 3 Nobel St., Moscow 143026, Russia}
\affiliation{Institute for Theoretical and Applied Electrodynamics, Russian
Academy of Sciences, Moscow, 125412, Russia}
\affiliation{Moscow Institute of Physics and Technology, Institutsky lane
9, Dolgoprudny, Moscow region, 141700, Russia}
   
\author{Alfred Zong}
\affiliation{Massachusetts Institute of Technology, Department of Physics,
Cambridge, Massachusetts 02139, USA}
   
\author{Anshul Kogar}
\affiliation{Massachusetts Institute of Technology, Department of Physics,
Cambridge, Massachusetts 02139, USA}  

\author{Nuh Gedik}
\affiliation{Massachusetts Institute of Technology, Department of Physics,
Cambridge, Massachusetts 02139, USA}

\author{Boris V. Fine}
\email{b.fine@skoltech.ru}
\affiliation{Skolkovo Institute of Science and Technology, Skolkovo
Innovation Center, 3 Nobel St., Moscow 143026, Russia}
\affiliation{Institute for Theoretical Physics, University of Heidelberg,
Philosophenweg 12, 69120 Heidelberg, Germany} 
   
\date{\today}

\begin{abstract}
We formulate a dynamical model to describe a photo-induced charge density
wave (CDW) quench transition and apply it to recent multi-probe
experiments on
LaTe$_3$
[A.~Zong \textit{et al.}, Nat. Phys. \textbf{15}, 27 (2019)].
Our approach relies on coupled time-dependent Ginzburg-Landau equations
tracking two order parameters that represent the modulations of the
electronic density and the ionic positions. We aim at describing
the amplitude of the order
parameters under the assumption that they are homogeneous in space.
This description is supplemented by a three-temperature model, which
treats separately the electronic temperature, temperature of the lattice
phonons with stronger couplings to the electronic subsystem, and
temperature of all other phonons. The broad scope of available data for
LaTe$_3$
and similar materials as well as the synergy between different
time-resolved spectroscopies allow us to extract model parameters. The 
resulting
calculations are in good agreement with ultra-fast electron diffraction
experiments, reproducing qualitative and quantitative features of the CDW
amplitude evolution during the initial few picoseconds after
photoexcitation.
\end{abstract}

\maketitle

\section{Introduction}

Dynamics of phase transitions associated with spontaneous symmetry breaking
remains an interesting subject both theoretically and experimentally.
Thanks to the advances in time-resolved pump-probe techniques, it is now
possible~[\!\!~\citenum{luo_pumpprobe_sc2010,chia_sdw_pumpprobe_exp2010,
yusupov_nat_phys2010,mihalovic_j_phys2013,mertelj_topol_defects2013,
zhang_pumpprobe_sc2013,suzuki_sdw_exp2017,Alfred2018,
naseska_sdw_pumpprobe_exp2018}]
to perturb an ordered state and then monitor its fast non-adiabatic
recovery. For strong perturbations, one can observe a passage through an
ordering transition, register the emergence of ordered phases, and measure
time evolution of diverse system parameters with a subpicosecond
resolution. The responses of ordered phases, such as superconducting
phase~[\!\!~\citenum{luo_pumpprobe_sc2010,zhang_pumpprobe_sc2013,
noneq_cooper_instba2016,higgs_sc2015}],
spin-density-wave~[\!\!~\citenum{chia_sdw_pumpprobe_exp2010,suzuki_sdw_exp2017,
naseska_sdw_pumpprobe_exp2018}]
and
charge-density-wave~[\!\!~\citenum{trARPES_tbte3_schmitt2008,yusupov_nat_phys2010,
Eichberger2010,
Hellmann2010,Rohwer2011,Petersen2011,Erasmus2012,mihalovic_j_phys2013,
mertelj_topol_defects2013,Rettig2016,Moore2016,Haupt2016,Vogelgesang2017,
Laulhe2017,Alfred2018,AlfredPavelPRL,sm_te3_exp_gl2019,Kogar2019,Zhou2019}]
phases, have been studied this way.

The focus of the present work is on the non-equilibrium dynamics across a
CDW transition. Despite long and thorough
scrutiny~[\!\!~\citenum{cdw_collection_gorkov_gruner,gruner_review_cdw1988}],
the CDW state continues to generate ample amount of research activity
motivated by interesting many-body physics (collective transport
phenomena~[\!\!~\citenum{gruner_review_cdw1988,bolloch_cdw_transport_exp2008,
monceau_cdw_review2012,pinsolle_prl_cdw_exp2012}],
non-mean field critical
exponents~[\!\!~\citenum{girault_cdw_crit_expon_exp1989,
ru_3Te_chem_pressure2008,hoesch_cdw_crit_expon_exp2009}],
exotic metastable `hidden' 
states~[\!\!~\citenum{stojchevska_hidden_cdw2014,yoshida_hidden_cdw2014,
vaskivskyi_hidden_cdw2016,karpov_model_sci_rep2018}]),
and large number of experimentally available model systems. In particular,
one can mention the  ``classical'' CDW materials, such
as~[\!\!~\citenum{nise3_cdw_exp1976,nbse2_cdw_exp1977,tas2_exp_cdw1984,
blue_bronzes_exp_cdw1983}]
NbSe$_3$,
NbSe$_2$,
TaS$_2$,
blue bronzes
K$_{0.30}$MoO$_3$
and
Rb$_{0.30}$MoO$_3$.
 CDW phase was also observed and actively investigated in
the family of rare-earth tritellurides
$R$Te$_3$~[\!\!~\citenum{dimasi_3te_cdw_exp1995,gweon_3te_cdw_exp1998,
laverock_nest_3Te2005,ru_tritell_transp_theromod2006,
lavagnini_3Te_phonons2008,ru_3Te_chem_pressure2008,Yusupov_PRL_2008,
yusupov_nat_phys2010,eiter_3Te_raman2013,mertelj_topol_defects2013,
mihalovic_j_phys2013,bulloch_late3_xray2016,hu_3Te_optics2014,
sinchenko_3Te_magnetores2017,sm_te3_exp_gl2019}].

Recently, we
reported~[\!\!~\citenum{Alfred2018}]
results of an experimental multi-probe study of a photo-induced CDW
transition in a member of the rare-earth tritelluride family
LaTe$_3$.
In these experiments, the post-pump relaxation was monitored with the help
of three different time-resolved techniques: ultra-fast electron
diffraction (UED), transient reflectivity, and time- and angle-resolved
photo-emission spectroscopy (tr-ARPES). These measurements delivered a
wealth of complementary information about both electronic and lattice
degrees of freedom, and, at the same time, highlighted the need for
advancing a quantitative theoretical description of the
far-from-equilibrium dynamics in CDW materials and beyond. In the present
paper, the experiments of
Ref.~\citenum{Alfred2018}
will serve as the primary testing ground for a rather general theoretical
approach.

The CDW order is characterized by the amplitude and the phase of charge
modulations. Both can strongly fluctuate in time and space. The amplitude
fluctuations are sometimes referred to as ``the Higgs modes", while the
phase fluctuations are associated with the appearance of phasons and
topological defects,  e.g., dislocations.  The experiments of
Ref.~\citenum{Alfred2018}
produced evidence  that the
relaxation of the phase is significantly slower than that of the amplitude.
The slowness of the phase relaxation was interpreted as being due to the
presence of topological defects. Modeling the latter, however, is rather
expensive computationally, because it requires space-resolved simulations.
It is, therefore, reasonable to ask first what part of the observed
phenomenology can be accounted for by the amplitude relaxation only. This
is what we do in the present work. The phase relaxation is to be
investigated elsewhere.

In what follows, we develop a theoretical framework for describing the time
evolution of {\it space-averaged} CDW amplitude in response to a strong
quench induced by a femtosecond laser pulse. In the course of the ensuing
nonequilibrium evolution, electrons relax much faster than the lattice.
Therefore, the modulation of the electronic density and the modulation of
the lattice should be treated as two distinct components of the CDW order.
We do this using time-dependent Ginzburg-Landau (TDGL)
equations~[\!\!~\citenum{mcmillan_tdgl1975,yusupov_nat_phys2010,
mihalovic_j_phys2013,Schafer_PRL_2010,Schaefer_PRB_2014,
schuller_gray_tdgl2006,
sm_te3_exp_gl2019,AlfredPavelPRL,period_doubling2014_nature}]
with two order
parameters~[\!\!~\citenum{Schafer_PRL_2010,Schaefer_PRB_2014}].
We also approximate energy redistribution between different degrees of
freedom using the so-called ``three-temperature
model"
[\!\!~\citenum{perfetti_3temp_mod2007,mansart_3temp_mod2010,
mansart_3temp2013,mansart_3temp_pnas2012,3-temp,review_ultrafast2017}],
which assigns separate temperatures to (i)~electrons, (ii)~phonons
efficiently coupled to the electronic subsystem, and (iii)~all other
phonons. Both TDGL equation with two order parameters and the three
temperature model were previously considered in the above-mentioned
references but not in combination with each other. The former was also
tested only in the small-oscillations
regime~[\!\!~\citenum{Schafer_PRL_2010,Schaefer_PRB_2014}],
which characterizes the response of the system to a weak quench.

Fusing together the two-order-parameter TDGL equations and the
three-temperature model into a single capable formalism is the main agenda
of this paper. Our analysis indicates that the combination of the above two
ingredients constitutes a quantitatively adequate and yet efficient
theoretical framework for treating {\it strong} photo-induced quenches in
CDW materials. This framework should also be applicable to other materials,
where electronic order is coupled to the lattice. 

The paper is organized as follows. In
Sec.~\ref{sec::formalism}
we present the theoretical formalism. The values of parameters for the
resulting equations are fixed in
Sec.~\ref{sec::choice_of_params}.
Numerical simulations are compared with experimental data in
Sec.~\ref{sec::experiment}.
Emergent time scales are summarized in
Sec.~\ref{sec::time_scales}.
Section~\ref{sec::discussion}
contains discussion. Finally, the conclusions are presented in
Sec.~\ref{sec::conclusions}.
Technically involved derivations are relegated to the Appendices.

\section{Theoretical formalism}
%%%%%%%%%%%%%%%%%%%%%%%%%%%%%%%%%%%%%%%%%%%%%%%%%%
\label{sec::formalism}
%%%%%%%%%%%%%%%%%%%%%%%%%%%%%%%%%%%%%%%%%%%%%%%%%% 

%A CDW state, by its very nature, depends crucially on electrons and
%lattice cooperation. As a result, a generic non-equilibrium evolution of a
%CDW system involves both electronic and lattice degrees of freedom coupled
%together into a complex kinetic ensemble.
\subsection{Preliminary qualitative considerations}
%%%%%%%%%%%%%%%%%%%%%%%%%%%%%%%%%%%%%%%%%%%%%%%%%% 
\label{preliminary}
%%%%%%%%%%%%%%%%%%%%%%%%%%%%%%%%%%%%%%%%%%%%%%%%%% 

In an experiment, a laser pumping pulse initially excites mostly electronic
degrees of freedom, while keeping the lattice unaffected. The ensuing
internal thermalization of the electronic subsystem is much faster than
that of the lattice. It occurs via electron-electron interactions. 
Given our assumption of a strong laser pulse, the resulting electronic
temperature is significantly higher than the initial temperature of the
system -- possibly higher than the CDW transition temperature. To be
specific, let us focus on the latter possibility. In such a case, if the
system were completely in equilibrium, then electronic density modulations
would quickly disappear. However, since the ions had no time to respond
yet, the lattice modulation associated with CDW remains intact, which
imposes an external periodic potential on the electronic subsystem.
Therefore, once the electronic density relaxes, the electronic order
parameter starts tracking the lattice order parameter. During the
subsequent evolution, the lattice, on the one hand, experiences a
diminished push from electrons to assume a modulated structure; hence the
amplitude of the lattice modulation gradually decreases. This happens in an
oscillatory way due to the motion of heavy ions near their equilibrium
positions. On the other hand, the electronic subsystem, whose heat capacity
is much smaller than that of the lattice, rapidly loses energy to the
lattice. As a result, the electronic temperature also decreases and
eventually falls below the temperature of the CDW phase transition. Once
this happens, both the electronic and the lattice orders start recovering,
while the electronic temperature continues to decrease until it reaches the
temperature of the lattice.

%At first sight, a model describing the above scenario requires only two
%temperatures, namely, those of electrons and the lattice. However, having
%attempted to use only two temperatures, we obtained rather noticeable
%discrepancies with experiments. Therefore, like a number of other
%authors[\!\!~\citenum{}, we arrived to the need to divide the lattice into two
%unequal subsystems and introduce separate temperatures for each of them.
%The smaller lattice subsystem represents phonons in the vicinity of the
%CDW wave vector. These phonons  strongly couple to the electron susbsystem
%and hence initially absorb most energy from the electronic subsystem.  The
%larger lattice subsystem represents all other lattice degrees of freedom,
%which then absorb energy from the smaller one. 
 
Below we develop a theoretical model capturing the above non-equilibrium
evolution. It describes the electronic and the lattice CDW amplitudes via
the TDGL formalism. For a non-equilibrium state, we employ the so-called
three temperature model, where the electronic and the two lattice
subsystems are are assumed to thermalize internally to quasi-equilibrium
distributions characterized by different temperatures. We then use rate
equations to describe the energy flow between these subsystems.

\subsection{TDGL sector}

Among theoretical
tools~[\!\!~\citenum{artemenko_volkov1981,artemenko_cdw2003,
cdw_collection_gorkov_gruner,takane2016,devereaux1,devereaux2,devereaux3,
devereaux4,stefanucci2013,aoki_rmp2014,murakami_sc2017}]
capturing the dynamics of an order parameter, the TDGL
equation~[\!\!~\citenum{mcmillan_tdgl1975,yusupov_nat_phys2010,
mihalovic_j_phys2013,Schafer_PRL_2010,Schaefer_PRB_2014,
schuller_gray_tdgl2006,sm_te3_exp_gl2019,AlfredPavelPRL,
period_doubling2014_nature}]
is one of the most popular. Despite the known issues with its microscopic
justification~[\!\!~\citenum{kopnin_book_tdgl2001}],
TDGL formalism remains in wide use due to its simplicity and intuitive
appeal. Below, we introduce the ingredients of this formalism adapted to a
setting with two order parameters.

\subsubsection{Static Landau functional}
%%%%%%%%%%%%%%%%%%%%%%%%%%%%%%%%%%%%%%%%%%%%%%%%%%
\label{subsec::landau_theory}
%%%%%%%%%%%%%%%%%%%%%%%%%%%%%%%%%%%%%%%%%%%%%%%%%% 

The CDW state is characterized by both the modulation of the electronic
density
\begin{eqnarray}
%%%%%%%%%%%%%%%%%%%%%%%%%%%%%%%%%%%%%%%%%%%%%%%%%%
\label{eq::elec_dens}
%%%%%%%%%%%%%%%%%%%%%%%%%%%%%%%%%%%%%%%%%%%%%%%%%% 
\delta \rho_{\rm e} ({\bf r})
=
A \exp (i{\bf Q} \cdot {\bf r})
+ {\rm c.c.},
\end{eqnarray}
and the displacements of ions
\begin{eqnarray}
\delta {\bf r}_{n} 
=
\left[ i u \exp (i{\bf Q} \cdot {\bf r}_n) + {\rm c.c.} \right]
\hat{e}_{\rm CDW}
\end{eqnarray}
from high-symmetry lattice positions
${\bf r}_{n}$.
Here
$\bf Q$
is the CDW wave vector;
$\hat{e}_{\rm CDW}$
is the unit vector along the CDW displacements. Quantities $A$ and $u$
describe complex electronic and lattice order parameters. In
Appendix~\ref{app::bragg_sum_rule},
we discuss the connection between the lattice order and the diffraction
measurements.

To describe equilibrium properties of the CDW system, one can introduce a
Landau-type
functional~[\!\!~\citenum{landau_staty}],
which depends either exclusively on $A$, or
exclusively on $u$, because in equilibrium $A$ is proportional to $u$ (for
sufficiently small $A$ and $u$). However, to account for the
non-equilibrium properties of the CDW during photo-induced transition, we
need both $A$ and $u$. We thus consider the Landau functional of the
following
form~[\!\!~\citenum{Schafer_PRL_2010,Schaefer_PRB_2014}]:
\begin{equation}
%%%%%%%%%%%%%%%%%%%%%%%%%%%%%%%%%%%%%%%%%%%%%%%%%%
\label{eq::landau_func}
%%%%%%%%%%%%%%%%%%%%%%%%%%%%%%%%%%%%%%%%%%%%%%%%%% 
{\cal F}[A,u] = - a |A|^2 + \frac{b}{2} |A|^4
- \eta(A u^* + A^* u ) + {\cal K} |u|^2,
\end{equation}
where $a$, $b$, $\eta$ and
${\cal K}$
are the expansion parameters. The first two terms in
Eq.~(\ref{eq::landau_func})
are of purely electronic origin. The last term corresponds to the elastic
lattice energy, which increases if the ions are shifted from their most
symmetric positions. Finally, the term proportional to $\eta$ describes the
electron-lattice coupling -- often the main driving force behind the CDW
transition. 

Below we assume that parameters of the Landau functional are temperature
independent, except for
\begin{eqnarray}
%%%%%%%%%%%%%%%%%%%%%%%%%%%%%%%%%%%%%%%%%%%%%%%%%%
\label{eq::a_def}
%%%%%%%%%%%%%%%%%%%%%%%%%%%%%%%%%%%%%%%%%%%%%%%%%% 
a = \alpha (T_0 - T),
\end{eqnarray} 
where $\alpha$ is a positive proportionality coefficient, $T$ is the
temperature, and
$T_0$
is the ``bare" transition temperature for a hypothetical situation of
vanishing electron-lattice interaction. (In principle, the parameter
$T_0$ 
is not a physical temperature but simply a parameter characterizing the structure
of the Landau functional; hence, it can be negative.) Due to finite
coupling between $A$ and $u$, the actual transition into the ordered phase
occurs at the critical temperature
\begin{eqnarray}
T_{\rm c} = T_0 + \frac{\eta^2}{\alpha{\cal K}}.
\end{eqnarray}
For
LaTe$_3$,
$T_{\rm c} \approx 670$K~[\!\!~\citenum{Alfred2018}].
As for
$T_0$,
it can be estimated as
\begin{equation}
%%%%%%%%%%%%%%%%%%%%%%%%%%%%%%%%%%%%%%%%%%%%%%%%%%
\label{eq::T0}
%%%%%%%%%%%%%%%%%%%%%%%%%%%%%%%%%%%%%%%%%%%%%%%%%%
T_0 = T_{\rm c} (1 - \zeta),
\end{equation}
where
\begin{equation}
\zeta = \frac{\eta^2}{\alpha  T_{\rm c}{\cal K}}
\end{equation}
characterizes the strength of the electron-phonon interaction. For the
parameters chosen in
Sec.~\ref{sec::choice_of_params}
to represent
LaTe$_3$,
we obtained
$T_0 = -67$\,K.

Minimizing the
functional~(\ref{eq::landau_func})
for
$T < T_{\rm c}$,
one finds equilibrium values of the order parameters:
\begin{eqnarray}
A_{\rm eq} = \sqrt{\frac{\alpha}{b} (T_{\rm c} - T)},
\quad
u_{\rm eq} = \frac{\eta}{{\cal K}} A_{\rm eq}.
\end{eqnarray}
For our calculations, it is convenient to work with the dimensionless
quantities:
\begin{eqnarray}
%%%%%%%%%%%%%%%%%%%%%%%%%%%%%%%%%%%%%%%%%%%%%%%%%% 
\label{dimensionless}
%%%%%%%%%%%%%%%%%%%%%%%%%%%%%%%%%%%%%%%%%%%%%%%%%% 
x = \frac{A}{A_{\rm eq}(T = 0)},
\quad
y = \frac{u}{u_{\rm eq}(T = 0)}.
\end{eqnarray}
For
$T \leq T_{\rm c}$,
equilibrium values of $x$ and $y$ are
\begin{eqnarray}
%%%%%%%%%%%%%%%%%%%%%%%%%%%%%%%%%%%%%%%%%%%%%%%%%% 
\label{eq::equillibrium}
%%%%%%%%%%%%%%%%%%%%%%%%%%%%%%%%%%%%%%%%%%%%%%%%%% 
x_{\rm eq} = y_{\rm eq} = \sqrt{{\Theta}},
\end{eqnarray}
where
${\Theta} = (T_{\rm c}-T)/T_{\rm c}$.

\subsubsection{Time-dependent equations}
%%%%%%%%%%%%%%%%%%%%%%%%%%%%%%%%%%%%%%%%%%%%%%%%%%
\label{subsec::tdgl}
%%%%%%%%%%%%%%%%%%%%%%%%%%%%%%%%%%%%%%%%%%%%%%%%%% 

The next step is to generalize the static Landau theory to non-equilibrium
situations. We will describe the dynamics of the electronic degree of
freedom
$A$ as
\begin{eqnarray}
%%%%%%%%%%%%%%%%%%%%%%%%%%%%%%%%%%%%%%%%%%%%%%%%%% 
\label{eq::A_relax}
%%%%%%%%%%%%%%%%%%%%%%%%%%%%%%%%%%%%%%%%%%%%%%%%%% 
\Gamma \frac{d A}{dt} = -\frac{\partial {\cal F}}{\partial A^*}
=
a A - b |A|^2 A + \eta u,
\end{eqnarray}
where
$\Gamma$ is a damping parameter. In dimensionless
variables~(\ref{dimensionless}),
Eq.~(\ref{eq::A_relax})
reads 
\begin{eqnarray} 
%%%%%%%%%%%%%%%%%%%%%%%%%%%%%%%%%%%%%%%%%%%%%%%%%%
\label{eq::TDGL1}
%%%%%%%%%%%%%%%%%%%%%%%%%%%%%%%%%%%%%%%%%%%%%%%%%% 
\tau_0 \frac{d x}{dt} - \Theta x + |x|^2 x + \zeta (x - y)
= 0,
\end{eqnarray} 
where
$\tau_0 =  \frac{\Gamma}{\alpha T_{\rm c}}$
is the electronic density relaxation time.

From the viewpoint of the true microscopic kinetics, 
Eq.~(\ref{eq::TDGL1})
is a crude approximation. However, for the purposes of the present work,
this approximation should be sufficient given that we are mainly interested
in the dynamics of variable $y$, which unfolds on the time scale much
longer than
$\tau_0$.
This longer time scale is associated with the motion of ions, which are
much heavier and, thus, much slower than electrons. It is only important
for us that, on fast time scale of 
$\tau_0$,
variable $x$ relaxes to the ``instantaneous equilibrium" value 
$\bar{x}$
given by the real-valued root of the equation
\begin{eqnarray}
\Theta \bar{x} - \bar{x}^3 + \zeta (y - \bar{x})
= 0.
\end{eqnarray}

To model the evolution of
$u$, we keep in mind that
$u$ is associated
with displacements of heavy ions, which can be viewed, approximately, as
classical objects. The forces acting on the ions are
\begin{eqnarray}
f_u = - \frac{d {\cal F}}{d u^*}
=
- {\cal K} u + \eta A.
%%%%%%%%%%%%%%%%%%%%%%%%%%%%%%%%%%%%%%%%%%%%%%%%%%
\label{eq::force_u}
%%%%%%%%%%%%%%%%%%%%%%%%%%%%%%%%%%%%%%%%%%%%%%%%%% 
\end{eqnarray} 
The term proportional to 
${\cal K}$
in
Eq.~(\ref{eq::force_u})
describes elastic force that pulls ions back to their high-symmetry
positions. The term proportional to
$\eta$ originates from the interaction
with the modulated electron
density~(\ref{eq::elec_dens}).
By adding damping, we arrive at a classical equation of motion for
$u$. In
the rescaled
variables~(\ref{dimensionless}),
it reads
\begin{equation}
%%%%%%%%%%%%%%%%%%%%%%%%%%%%%%%%%%%%%%%%%%%%%%%%%%
\label{eq::TDGL2}
%%%%%%%%%%%%%%%%%%%%%%%%%%%%%%%%%%%%%%%%%%%%%%%%%% 
\frac{1}{\omega_0^2} \frac{d^2 y}{dt^2}
+
\frac{\gamma_y}{\omega_0} \frac{d y}{dt} + (y - x) = 0,
\end{equation}
where 
$\omega_0 = \sqrt{{\cal K}/ m}$
is unrenormalized frequency of a phonon mode actively involved in the CDW
formation (we call this mode ``the CDW phonon"); $m$ is the ion mass
parameter;
$\gamma_y$
describes damping.
Equations~(\ref{eq::TDGL1})
and~(\ref{eq::TDGL2})
constitute the desired TDGL equations in a dimensionless form.

Previously TDGL 
formalism~[\citenum{Schafer_PRL_2010,Schaefer_PRB_2014}]
with two order parameters was applied to describe small deviations from
equilibrium in a CDW material. However, in the present work we deal with a
far-from-equilibrium response, which requires an additional ingredient in
the theory. This additional ingredient is the three-temperature model
introduced in the next subsection.

\subsection{Three-temperature model}

Temperature $T$ that appears in
Eq.~(\ref{eq::a_def})
is the electronic temperature which, from now on, we denote as 
$T_{\rm e}$.
We need this change of notation to distinguish the electronic temperature 
from the temperature (or temperatures) of the lattice. The TDGL
equation~(\ref{eq::TDGL1})
thus depends on 
$T_{\rm e}$.

Let us expand on the qualitative scenario given in
Section~\ref{preliminary}
using the numbers specific for the experiment of
Ref.~\citenum{Alfred2018}.  
According to that scenario, all photons from the pumping laser pulse are
mostly absorbed by the electronic subsystem. Right after the laser pulse,
the electronic subsystem is far from equilibrium. However, it quickly
reaches quasistationary thermal state with temperature
$T_{\rm e} (0) \sim 1000$\,K,
which significantly exceeds the initial lattice temperature of about
300\,K.
Lattice phonons, whose heat capacity is much larger than that of the
electrons, then act as a heat sink gradually absorbing energy of the hot
electrons. Let us emphasize that, during this process, it can happen that
phonon distribution function becomes highly non-thermal.

The simplest approach to capture the above dynamics would be to introduce a
two-temperature
model~[\!\!~\citenum{kaganov_2temp_model1957,anisimov1974,
chen_2temp_model2005}]
tracking the electronic and the lattice temperatures. However, given all
available experimental data -- including (i)~heat capacity
measurements~[\!\!~\citenum{ru_tritell_transp_theromod2006}]
in
LaTe$_3$,
(ii)~UED data (in particular, the time dependence of the crystal Bragg
peaks
intensity~[\!\!~\citenum{Alfred2018}]),
(iii)~short-time transient-reflectivity
dynamics~[\!\!~\citenum{Alfred2018}],
and (iv)~tr-ARPES
data~[\!\!~\citenum{Alfred2018}],
which allows to estimate electronic heat capacity (see
Sec.~\ref{sec::choice_of_params}
below) -- we were unable to adequately reproduce the observed behavior
using the two-temperature model. 

We attribute the above discrepancy to an intrinsic limitation of the
two-temperature model. Namely, such a model  describes the entire lattice by a
single temperature, while, in reality, not all phonons are thermalized simultaneously.
Instead, we expect that the electronic energy
is preferentially absorbed by a smaller subset of the phonon ensemble. Such a
selective coupling was discussed for rare-earth tritellurides in
Refs.~\citenum{tbte3_phonon_maschek_exp2015,
dyte3_phonon_maschek_exp2018}
and, in a broader context, in
Refs.~\citenum{mansart_3temp_pnas2012,h_nbs2_electr_phon_coupl_anisotr2012,
nonuniform_electron_phon_coupl_nbse2,review_ultrafast2017}.
In particular, the authors of
Ref.~\citenum{mansart_3temp2013}
argued that layered structure makes preferential electron coupling to a
subset of phonon modes more likely. For
LaTe$_3$,
we can identify two factors that increase the selectivity of the
electron-phonon coupling. First, electronic states forming the Fermi
surface are located in the tellurium-only 
layers~\cite{dimasi_3te_cdw_exp1995}
and, hence, are weakly coupled to the lattice degrees of freedom localized
in the 
$R$Te
layers. Second, at sufficiently low excitation energies of the electronic
quasiparticles, the momentum conservation law implies that the probability 
of emission (or absorption) of a phonon with momentum 
${\bf q}$
is proportional to
$\sum_{\bf k} \delta ( \varepsilon_{\bf k} - \varepsilon_{\rm F})
\delta ( \varepsilon_{{\bf k} + {\bf q}} - \varepsilon_{\rm F})$,
where
$\varepsilon_{\rm F}$
is the Fermi energy, and
$\varepsilon_{\bf k}$
is the electron dispersion. In the rare-earth tritellurides, due to
proximity to the Fermi surface nesting condition, the latter sum is a
strongly non-uniform function of the momentum 
${\bf q}$,
see Eq.~(2) and Fig.~6a of
Ref.~\citenum{johannes_mazin}.
This non-uniformity imposes additional selection criteria on the phonon
modes participating in the electron scattering.

In order to account for the selectivity of the electron-phonon coupling, we
split the lattice phonons into two subgroups: (i)~the ``hot phonons'', the
ones strongly coupled to the electrons, and (ii)~``cold phonons'', i.e. the
rest of the lattice modes. The hot phonons are to be characterized by
temperature
$T_{\rm L2}$,
while cold phonons by temperature
$T_{\rm L1}$.
To simplify the model, we neglect the direct energy transfer between the
electronic subsystem and the cold phonons. 

The post-pulse temperature dynamics is then described by the
three-temperature
model~[\!\!~\citenum{perfetti_3temp_mod2007,mansart_3temp_mod2010}]:
\begin{eqnarray}
%%%%%%%%%%%%%%%%%%%%%%%%%%%%%%%%%%%%%%%%%%%%%%%%%%
\label{eq::t3m1}
%%%%%%%%%%%%%%%%%%%%%%%%%%%%%%%%%%%%%%%%%%%%%%%%%% 
&&C_{\rm e}(T_{\rm e})
\frac{d T_{\rm e}}{d t}
=
- G_{\rm eL}(T_{\rm e} - T_{\rm L2}),
\\
%%%%%%%%%%%%%%%%%%%%%%%%%%%%%%%%%%%%%%%%%%%%%%%%%%
\label{eq::t3m2}
%%%%%%%%%%%%%%%%%%%%%%%%%%%%%%%%%%%%%%%%%%%%%%%%%% 
&&C_{\rm L1}\displaystyle \frac{d T_{\rm L1}}{dt}
=
- G_{\rm LL}( T_{\rm L1} -T_{\rm L2}),
\\
%%%%%%%%%%%%%%%%%%%%%%%%%%%%%%%%%%%%%%%%%%%%%%%%%%
\label{eq::t3m3}
%%%%%%%%%%%%%%%%%%%%%%%%%%%%%%%%%%%%%%%%%%%%%%%%%% 
&&C_{\rm L2} \frac{d T_{\rm L2}}{dt} 
=
- G_{\rm eL}( T_{\rm L2} -T_{\rm e})
- G_{\rm LL}( T_{\rm L2} -T_{\rm L1}).
\quad
\end{eqnarray}
Here
$C_{\rm L1}$
and
$C_{\rm L2}$
are the heat capacities of the cold phonons and of the hot phonons,
respectively;
$C_{\rm e} (T_{\rm e})$
is the temperature-dependent electronic heat capacity whose functional form
is specified in
subsection~\ref{subsec::app_choice_params_3temp}.
Parameters
$G_{\rm eL}$
and
$G_{\rm LL}$
describe energy exchange rates.

The initial conditions for 
Eqs.~(\ref{eq::t3m1}-\ref{eq::t3m3})
are to be chosen as follows. For the lattice,
$T_{\rm L1} (0) = T_{\rm L2} (0) = T_{\rm env}$,
where
$T_{\rm env}
$,
is the pre-pulse equilibrium temperature of all three subsystems. For
electrons,
$T_{\rm e} (0)$
is defined as the temperature right after the laser pulse and the ensuing
fast electronic self-thermalization. The value
$T_{\rm e} (0)$
is, therefore, a function of absorbed electromagnetic energy per mole,
which is, in turn, proportional to (i)~the photoexcitation density
$F$ (the
number of absorbed photons per unit volume), (ii)~the pump photon energy
$\hbar \omega_\gamma
$,
and (iii)~the molar volume 
${\cal V}
$
of
LaTe$_3$.
The energy balance condition then gives the equation determining 
$T_{\rm e} (0)$:
\begin{eqnarray}
%%%%%%%%%%%%%%%%%%%%%%%%%%%%%%%%%%%%%%%%%%%%%%%%%%
\label{eq::init_Te}
%%%%%%%%%%%%%%%%%%%%%%%%%%%%%%%%%%%%%%%%%%%%%%%%%% 
\hbar \omega_\gamma {\cal V} F
=
\int\limits_{T_{\rm env}}^{T_{\rm e}(0)} C_{\rm e} (T) dT.
\end{eqnarray}

\section{Choice of parameters for $\text{LaTe}_3$}
%%%%%%%%%%%%%%%%%%%%%%%%%%%%%%%%%%%%%%%%%%%%%%%%%%
\label{sec::choice_of_params}
%%%%%%%%%%%%%%%%%%%%%%%%%%%%%%%%%%%%%%%%%%%%%%%%%% 

Overall, our formalism includes five
equations:~(\ref{eq::TDGL1})
and~(\ref{eq::TDGL2})
for the TDGL sector, 
and~(\ref{eq::t3m1}--\ref{eq::t3m3})
for the temperature-evolution sector. To perform simulations, one needs to
select specific values for the model parameters: 
$\omega_0$,
$\tau_0$,
$\zeta$,
and
$\gamma_y$
in the TDGL sector; and
$C_{\rm L1}$,
$C_{\rm L2}$,
$G_{\rm eL}$,
and
$G_{\rm LL}$
in the temperature-evolution sector. One also needs a concrete functional
form of the temperature dependence of the electronic heat capacity
$C_{\rm e} (T_{\rm e})$.

Since the total number of free parameters is large, extracting their values
through an indiscriminate fitting might produce misleading results. To
circumvent this issue, we split the whole task into several steps to be
presented in the following subsections. In each step, only a small number
of the unknowns are fixed. This approach relies on the availability of a
broad array of experimental results for the rare-earth tritellurides
$R$Te$_3$,
in particular 
LaTe$_3$~[\!\!~\citenum{Alfred2018}].
For convenience of the readers, the final values of all parameters are
collected in 
Table~\ref{tab::params_vals}.
\begin{table*}[t!]
  \begin{center}
    \begin{tabular}{||c|c|c||c|c|c||} 
      \hline
      \hline
	\multicolumn{3}{||c||}{TDGL sector} & 
	\multicolumn{3}{c||}{Temperature-evolution sector} \\
	\hline
{Parameter} & {Value} & Physical meaning & {Parameter} & {Value} & Physical
meaning \\
      \hline
      \hline
$\omega_0/(2\pi)$ & 3.1\,THz & Unrenormalized frequency &
$c_0$ & 1.1 mJ/mol$\cdot$K$^2$ & Coefficient in the low-temperature \\
& & of the CDW phonon & & &
electronic heat capacity, Eq.~(\ref{eq::electr_heat_cap0}) \\
      \hline
$\tau_0$ & 20\,{fs}   &  Electronic density &
$c$ & 4 mJ/mol$\cdot$K$^2$ & Coefficient in the high-temperature \\
& & relaxation time, Eq.~(\ref{eq::TDGL2}) & & &
electronic heat capacity, Eq.~(\ref{eq::electr_heat_cap}) \\
      \hline
$\zeta$ & 1.1 & Critical temperature renormalization & 
$C_{\rm tot}$& 99.7\,J/mol$\cdot$K & Total heat capacity of the lattice\\
& & parameter, Eq.~(\ref{eq::T0}) & & &\\
      \hline
$\gamma_y$ & $0.04$ & Damping parameter for &
$\kappa$ & 0.2 & Fraction of ``hot phonons" with \\
 & & the CDW phonon & & & respect to all phonons\\
      \hline
	$T_{\rm c}$ & 670 K & Temperature of the CDW &
$G_{\rm eL}$ & 5.5 J/ps$\cdot$K$\cdot$mol & Energy exchange rate between \\
 & & phase transition & & & electrons and ``hot phonons" \\
      \hline
 & & &
$G_{\rm LL}$ & 7.25 J/ps$\cdot$K$\cdot$mol & Energy exchange rate
between ``hot \\
 & & & & & phonons" and the rest of the lattice \\
     \hline
     & & &
$\tau_{\rm DW}$ & 2.2 ps & Time constant for the second stage\\
 & & & & & of the temperature relaxation, Eq.~(\ref{eq::tau_DW}) \\
\hline
 & & & $T_{\rm env}$ & 300K & Equilibrium temperature of the entire \\
 & & & & & system before the laser pulse \\
      \hline
      \hline
    \end{tabular}
  \end{center}
\caption{Parameters used for numerical simulations of
Eqs.~(\ref{eq::TDGL1})
and~(\ref{eq::TDGL2}),
and
Eqs.~(\ref{eq::t3m1}--\ref{eq::t3m3}).
%%%%%%%%%%%%%%%%%%%%%%%%%%%%%%%%%%%%%%%%%%%%%%%%%% 
\label{tab::params_vals}
%%%%%%%%%%%%%%%%%%%%%%%%%%%%%%%%%%%%%%%%%%%%%%%%%% 
}
\end{table*}

\subsection{TDGL parameters}
%%%%%%%%%%%%%%%%%%%%%%%%%%%%%%%%%%%%%%%%%%%%%%%%%%
\label{subsec::tdgl_params}
%%%%%%%%%%%%%%%%%%%%%%%%%%%%%%%%%%%%%%%%%%%%%%%%%% 

Here we fix the TDGL parameters by matching the frequency
$\omega_{\rm AM}$
and the damping constant 
$\gamma_{\rm AM}$
of the CDW amplitude mode (AM) obtained theoretically with the values
measured experimentally. Theoretically, we apply small-oscillations
formalism~[\!\!~\citenum{Schafer_PRL_2010,Schaefer_PRB_2014}]
to the TDGL
equations~(\ref{eq::TDGL1})
and~(\ref{eq::TDGL2}),
see
Appendix~\ref{app::small_oscillations}
for detailed derivations. On the experimental side, the phonon spectrum in
$R$Te$_3$
and its temperature dependence were reported in many
works~[\!\!~\citenum{lavagnini_3Te_phonons2008,eiter_3Te_raman2013,
tbte3_phonon_maschek_exp2015,dyte3_phonon_maschek_exp2018,Alfred2018}].

From the average frequency and the frequency width of the measured
transient reflectivity oscillations, we know
that~[\!\!~\citenum{Alfred2018}],
at 
$T=300$\,K,
\begin{eqnarray}
%%%%%%%%%%%%%%%%%%%%%%%%%%%%%%%%%%%%%%%%%%%%%%%%%%
\label{eq::AM_param}
%%%%%%%%%%%%%%%%%%%%%%%%%%%%%%%%%%%%%%%%%%%%%%%%%% 
\frac{\omega_{\rm AM}}{2 \pi} = \nu_{\rm AM} = 2.2\,{\rm THz},
\ 
\gamma_{\rm AM} = 1.26\times 10^{12}\,{\rm s}^{-1}.
\end{eqnarray}
We use  the above values to define a complex parameter
$\lambda = - \gamma_{\rm AM} + i \omega_{\rm AM}$,
which is  then substituted into
Eqs.~(\ref{eq::P_def}, \ref{eq::freq})
describing the small-oscillations eigenvalue problem. This leads to two
constraints on parameters
$\omega_0$,
$\tau_0$,
$\zeta$,
and
$\gamma_y$  expressed by
Eqs.~(\ref{eq::TDGL_params_constraints1})
and~(\ref{eq::TDGL_params_constraints2}).
Thereby, we reduce the total number of adjustable TDGL parameters from four
to two. For our analysis, it is convenient to treat
Eqs.~(\ref{eq::TDGL_params_constraints1})
and~(\ref{eq::TDGL_params_constraints2})
as definitions of the implicit functions
$\gamma_y (\omega_0, \tau_0)$
and
$\zeta (\omega_0, \tau_0)$,
which specify the dependence of $\zeta$ and
$\gamma_y$
on
$\omega_0$
and
$\tau_0$.
Thus, once 
$\omega_0$
and
$\tau_0$
are fixed, the TDGL sector contains no unknown parameters.

\begin{figure}[t!]
\includegraphics[width=9cm]{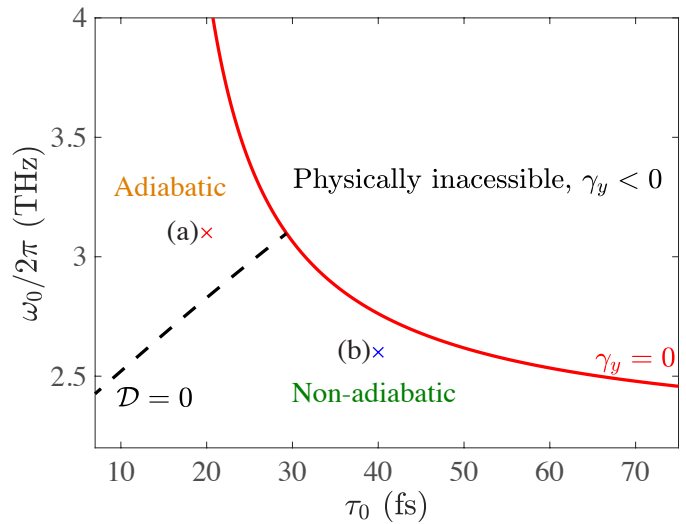}
\caption{Constraints on the allowed values of the TDGL parameters
$\tau_0$
and
$\omega_0$
formulated in
subsection~\ref{subsec::tdgl_params}.
The area above the solid red curve is physically inaccessible because it
corresponds to
$\gamma_y(\omega_0,\tau_0)<0$.
The available parameter space to the left and below the solid curve hosts
two regimes of small oscillations, adiabatic (above the dashed line) and
non-adiabatic (below the dashed line), see
subsection~\ref{subsec::tdgl_params}
and
Appendix~\ref{app::small_oscillations}.
These two regimes are exemplified by two points, (a) and (b), for which the
mode softening is illustrated in
Fig.~\ref{fig::frequencies}.
The dashed line is determined by equation
${\cal D} = 0$,
where 
${\cal D}$
is defined by
Eq.~(\ref{eq::D_def}).
The parameters
$\omega_0$
and
$\tau_0$
for
LaTe$_3$
are assumed to represent the adiabatic regime.
}
%%%%%%%%%%%%%%%%%%%%%%%%%%%%%%%%%%%%%%%%%%%%%%%%%% 
\label{fig::Phase_diag}
%%%%%%%%%%%%%%%%%%%%%%%%%%%%%%%%%%%%%%%%%%%%%%%%%%
\end{figure}

Estimating
$\omega_0$
and
$\tau_0$,
one must be mindful of several relevant theoretical and experimental
restrictions. The first of them is the physical requirement
$\gamma_y (\omega_0, \tau_0) \geq 0$.
It limits the allowed space for
$\omega_0$
and
$\tau_0$
to the region left and below the red curve in
Fig.~\ref{fig::Phase_diag}.

The next restriction is related to whether the AM frequency
$\omega_{\rm AM}$
softens to zero close to the CDW transition temperature
$T_{\rm c}$
or not. In
Refs.~\citenum{Schafer_PRL_2010,Schaefer_PRB_2014},
the former regime is called ``adiabatic'' and the latter ``non-adiabatic''
(see examples in
Appendix~\ref{app::small_oscillations}).
Although the behavior of the AM in
LaTe$_3$
near
$T_{\rm c}$
is not accessible experimentally, we rely on the reported universality of
the AM characteristics for several members of the
$R$Te$_3$
family~[\!\!~\citenum{Yusupov_PRL_2008}].
(The most noticeable aspect of this universality is the same
low-temperature value of the AM frequency
$\nu_{\rm AM} \approx 2.2$\,THz.)
Specifically, experiments suggest that the AM in 
TbTe$_3$
[see
Fig.~3(b,c) in
Ref.~\citenum{tbte3_phonon_maschek_exp2015}]
and DyTe$_3$
[see
Fig.~8(a,c) in
Ref.~\citenum{dyte3_phonon_maschek_exp2018}]
softens to zero close to the transition temperature. Thus, we assume the
adiabatic regime for
LaTe$_3$
as well. As shown in
Fig.~\ref{fig::Phase_diag},
such an assumption further confines 
$\omega_0$
and
$\tau_0$
to the region above the dashed line. Together with the previous constraint,
this implies that
$\tau_0 \lesssim 30$\,fs.

Now we note that the Heisenberg uncertainty principle suggests that
\begin{eqnarray}
%%%%%%%%%%%%%%%%%%%%%%%%%%%%%%%%%%%%%%%%%%%%%%%%%%
\label{eq::tau0_lower}
%%%%%%%%%%%%%%%%%%%%%%%%%%%%%%%%%%%%%%%%%%%%%%%%%%
\tau_0 \gtrsim \frac{1}{2\Delta} \approx 1\,{\rm fs},
\end{eqnarray} 
where
$2\Delta \approx 700$\,meV
is the CDW gap at
300\,K~[\!\!~\citenum{hu_3Te_optics2014}]. 
Given the above constraints, we assign
$\tau_0 = 20$\,{\rm fs}.
%This choice is in reasonable agreement with other estimates of purely
%electronic time scales. For example,
%Ref.~\citenum{mansart_3temp_pnas2012}
%concludes that in 
%Lu$_5$Ir$_4$Si$_{10}$
%the electronic thermalization time is
%$\sim 50$\,fs.
The dynamics of the CDW order parameters obtained from our numerical simulations is not very
sensitive to the specific choice of
$\tau_0$ as long as we are interested in  time scales much longer than $\tau_0$.

Finally, to set
$\omega_0$,
we use the relation
\begin{eqnarray}
%%%%%%%%%%%%%%%%%%%%%%%%%%%%%%%%%%%%%%%%%%%%%%%%%%
\label{eq::omega0}
%%%%%%%%%%%%%%%%%%%%%%%%%%%%%%%%%%%%%%%%%%%%%%%%%% 
\omega_{\rm AM} (T=0) = \lambda_{\rm CDW}^{1/2} \omega_0,
\end{eqnarray} 
where
$\lambda_{\rm CDW}$
is the electron-phonon coupling
constant~[\!\!~\citenum{gruner_book}]
responsible for the CDW instability. It is assumed both for the rare-earth 
tritellurides~[\!\citenum{eiter_3Te_raman2013}]
and for a broader class of materials [see, e.g., discussion after
Eq.~(5) in
Ref.~\citenum{cdw_collection_gorkov_gruner},
and Table~3.1 of
Ref.~\citenum{gruner_book}] and 
that
$\lambda_{\rm CDW} \approx 0.5$. 
When this value together with
$\omega_{\rm AM}$ given by
Eq.~(\ref{eq::AM_param}) 
is substituted into
Eq.~(\ref{eq::omega0}),
we obtain an estimate
$\omega_0/(2\pi) \approx 3.1$\,THz.
Here we assumed that
$\omega_{\rm AM}$
is the same at
$T=0$
and at
$T=T_{\rm env}$. 
Indeed, since 
$T_{\rm env}$ 
is significantly below
$T_c$,
it is permissible to treat
$\omega_{\rm AM}$
as being temperature-independent in the range
$T \leq T_{\rm env}$.

Once 
$\tau_0$
and
$\omega_0$
are determined, both
$\gamma_y = \gamma_y (\omega_0, \tau_0)$
and
$\zeta = \zeta (\omega_0, \tau_0)$
are obtained. The final values are summarized in
Table~\ref{tab::params_vals}.

\subsection{Three-temperature model parameters}
%%%%%%%%%%%%%%%%%%%%%%%%%%%%%%%%%%%%%%%%%%%%%%%%%%
\label{subsec::app_choice_params_3temp}
%%%%%%%%%%%%%%%%%%%%%%%%%%%%%%%%%%%%%%%%%%%%%%%%%% 

In the context of 
LaTe$_3$
experiments of
Ref.~\citenum{Alfred2018},
we assume that the equilibrium pre-pulse temperature of all three
subsystems is
$T_{\rm env} = 300$\,K.
For each value of the photoexcitation density $F$, the electronic
temperature right after the initial  self-thermalization of the electronic
subsystem
$T_{\rm e}(0)$
is to be calculated using
Eq.~(\ref{eq::init_Te})
with 
$\hbar \omega_\gamma = 1.19\,{\rm eV} = 1.9\times 10^{-19}\,{\rm J}$
and
${\cal V} = 76.8\,{\rm cm^3/mol}$.

Let us now turn to the temperature-dependence of
$C_{\rm e}(T_{\rm e})$.
At sufficiently low temperatures, electronic heat capacity is a linear
function of temperature 
\begin{eqnarray}
%%%%%%%%%%%%%%%%%%%%%%%%%%%%%%%%%%%%%%%%%%%%%%%%%
\label{eq::electr_heat_cap0}
%%%%%%%%%%%%%%%%%%%%%%%%%%%%%%%%%%%%%%%%%%%%%%%%% 
C^0_{\rm e} (T_{\rm e})
= c_0 T_{\rm e},
\end{eqnarray}
where~[\!\!~\citenum{ru_tritell_transp_theromod2006}]
$c_0 = 1.1$\,mJ/mol\,K$^2$
for
LaTe$_3$.
In the proximity to and above 
$T_{\rm c} = 670$\,K,
we do not expect 
Eq.~(\ref{eq::electr_heat_cap0}) 
to remain valid. It would imply that the electronic temperature following
the maximum intensity laser pulse reaches the value 
$\approx 4000$\,K,
while the analysis of our tr-ARPES data presented in
Appendix~\ref{app_subsect::tr_ARPES}
reveals that
$T_{\rm e} \lesssim 2000$\,K.
%see Fig.~\ref{fig:ARPES}.
The deviation from the linear temperature
dependence~(\ref{eq::electr_heat_cap0})
is also expected on the basis of purely theoretical reasoning outlined in
Appendix~\ref{app_subsect::pseudogap_heat_cap}.
Following that reasoning, we approximate 
$C_{\rm e}(T_{\rm e})$
by a piecewise linear ansatz:
\begin{eqnarray} 
%%%%%%%%%%%%%%%%%%%%%%%%%%%%%%%%%%%%%%%%%%%%%%%%%%
\label{eq::electr_heat_cap}
%%%%%%%%%%%%%%%%%%%%%%%%%%%%%%%%%%%%%%%%%%%%%%%%%% 
C_{\rm e}(T_{\rm e})\!
=\!
\begin{cases}
c_0 T_{\rm e}, & \!\!\!\!\!\text{if   } T_{\rm e} < T_{\rm env}, \\
c_0 T_{\rm env}
+
(T_{\rm e} - T_{\rm env}) c,\ & 
\!\!\!\!\!\text{if   } T_{\rm env} < T_{\rm e}, 
\end{cases}
\end{eqnarray}
where 
$c = 4$\,mJ$/$mol\,K$^{2}$
is a parameter extracted  from tr-ARPES experiments in
Appendix~\ref{app_subsect::pseudogap_heat_cap} .

\begin{figure}[t!]
\includegraphics[width=9cm]{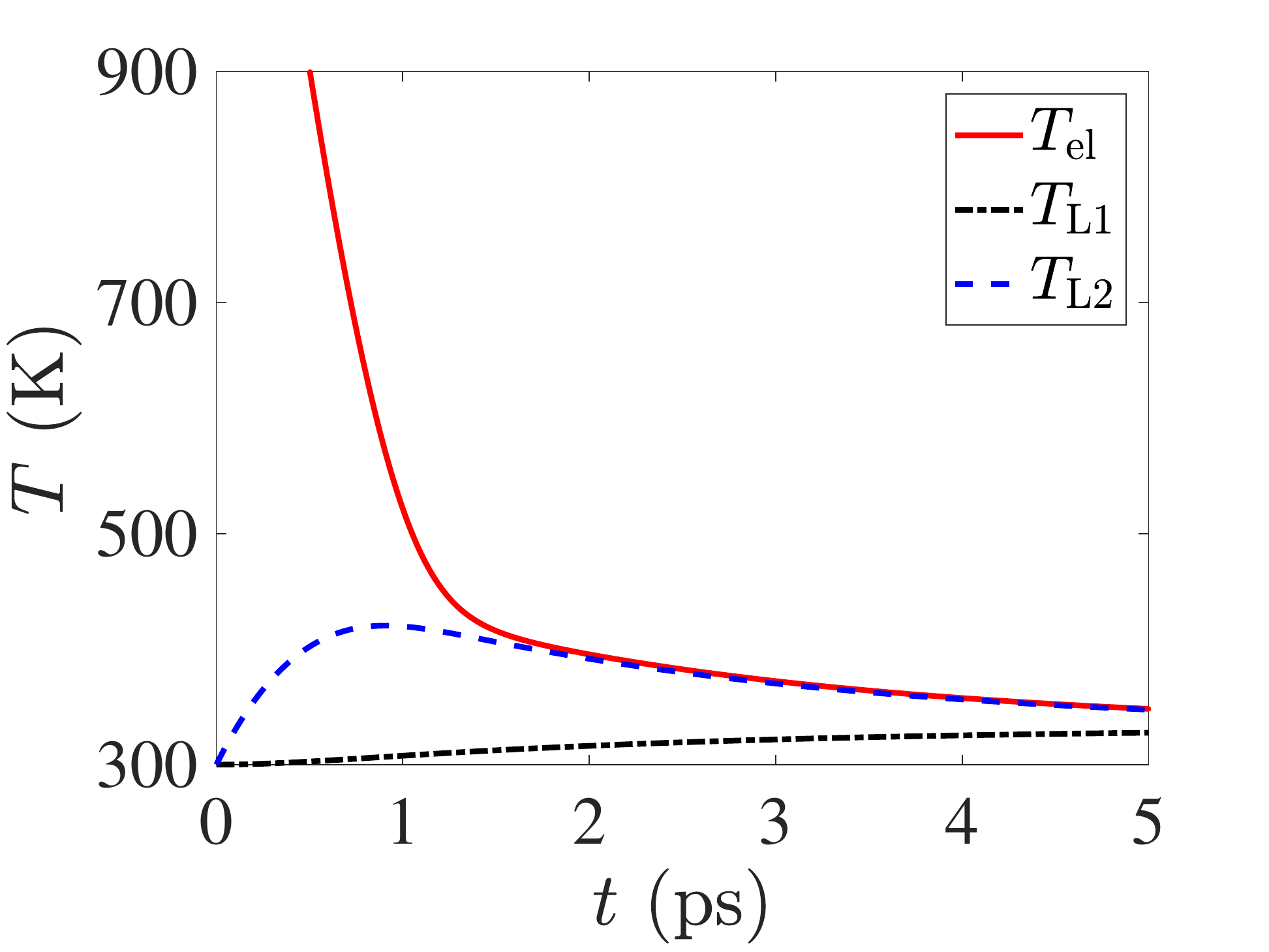}
\caption{Typical example of the time evolution of electronic temperature 
$T_{\rm e}$
(solid red line), temperature of hot phonons
$T_{\rm L2}$
(dashed blue line), and temperature of cold phonons
$T_{\rm L1}$
(dash-dotted black line). The curves are computed numerically on the basis
of
Eqs.~(\ref{eq::t3m1}-\ref{eq::t3m3})
with the initial value of 
$T_{\rm e}$
corresponding to the photoexcitation density
$F=2\times 10^{20}$\,cm$^{-3}$
and with parameters given in 
Table~\ref{tab::params_vals}.
%%%%%%%%%%%%%%%%%%%%%%%%%%%%%%%%%%%%%%%%%%%%%%%%%% 
\label{fig::3temp_evolution}
%%%%%%%%%%%%%%%%%%%%%%%%%%%%%%%%%%%%%%%%%%%%%%%%%% 
}
\end{figure}

Next we turn to lattice heat capacities
$C_{\rm L1}$
and
$C_{\rm L2}$.
Since these parameters are associated with two complementary groups of
phonons, we express
\begin{eqnarray}
C_{\rm L1} = (1 - \kappa) C_{\rm tot},\ 
C_{\rm L2} = \kappa C_{\rm tot},
\end{eqnarray}
where
$C_{\rm tot}$
is the total heat capacity of the lattice, and $\kappa$ is the coefficient
determining the fraction of the phonon modes contributing to
$C_{\rm L2}$.
We approximate 
$C_{\rm tot}$
by the Dulong-Petit value
$99.7$\,J/(mol\,K)
for
LaTe$_3$,
which is permissible in the temperature range of interest and consistent
with 
experiment~[\!\!~\citenum{ru_tritell_transp_theromod2006}].
The value of $\kappa$ is fixed to be equal to
$0.2$ in
subsection~\ref{subsect::Bragg_peak}
on the basis of our model fitting to the UED Bragg peak intensities.

A typical post-pulse time evolution of the three temperatures calculated on
basis of 
Eqs.~(\ref{eq::t3m1}-\ref{eq::t3m3})
is shown in
Fig.~\ref{fig::3temp_evolution}.
Here we assume that the initial rise of the electronic temperature occurs
on a very short time scale, which we approximate as instantaneous. The
remaining evolution can be divided into two stages. During the first stage,
$T_{\rm e}$
relaxes to
$T_{\rm L2}$
on the time scale of
1--2\,ps.
The second stage unfolds for
$t \gtrsim 1.5$\,ps,
where the common temperature of the electrons and the hot phonons
($T_{\rm e} \approx T_{\rm L2}$)
approaches
$T_{\rm L1}$.

For sufficiently strong laser pulses, such that 
$T_{\rm e} (0) \gg T_{\rm L1,L2}$,
the first stage can be accurately described by the approximate equation
\begin{eqnarray}
%%%%%%%%%%%%%%%%%%%%%%%%%%%%%%%%%%%%%%%%%%%%%%%%%%
\label{eq::Te_first_stage_approx}
%%%%%%%%%%%%%%%%%%%%%%%%%%%%%%%%%%%%%%%%%%%%%%%%%% 
C_{\rm e}(T_{\rm e})
\frac{d T_{\rm e}}{d t}
\approx
- G_{\rm eL}T_{\rm e},\label{eq::modified_1tm}
\end{eqnarray}
which is governed by a single parameter
$G_{\rm eL}$.
Its value can be estimated by assuming that, at high excitation densities,
the initial decay of the transient reflectivity, measured in
Ref.~\citenum{Alfred2018},
is controlled by 
$T_{\rm e} (t)$.
This way, we obtain
$G_{\rm eL} = 5.5$\,J/(ps$\cdot$K$\cdot$mol),
see
subsection~\ref{subsec::exp::reflectivity}
for further details.

During the second stage, the temperature relaxation process is exponential,
characterized by the time constant
\begin{eqnarray}
%%%%%%%%%%%%%%%%%%%%%%%%%%%%%%%%%%%%%%%%%%%%%%%%%% 
\label{eq::tau_DW}
%%%%%%%%%%%%%%%%%%%%%%%%%%%%%%%%%%%%%%%%%%%%%%%%%% 
\tau_{\rm DW} = \kappa(1-\kappa) C_{\rm tot} /G_{\rm LL}.
\end{eqnarray}
Here, following the notation of
Ref.~\citenum{Alfred2018},
we use the subscript `DW', which stands for `Debye-Waller', because the
above time constant controls the  evolution of the Bragg peak intensity in
the late-time regime.
Expression~(\ref{eq::tau_DW})
can be derived with the help of
Eqs.~(\ref{eq::t3m2})
and~(\ref{eq::t3m3})
in the limit
$T_{\rm e} = T_{\rm L2}$
(corresponding to
$t > 1.5$\,ps
in 
Fig.~\ref{fig::3temp_evolution}).
From the measured relaxation of the Bragg peak
intensity~[\!\!~\citenum{Alfred2018}],
we have
$\tau_{\rm DW} = 2.2$\,ps.
Thereby, 
Eq.~(\ref{eq::tau_DW})
defines
$G_{\rm LL} = 7.25$\,J/(ps$\cdot$K$\cdot$mol).

\section{Comparison of the experiments with the numerical simulations}
%%%%%%%%%%%%%%%%%%%%%%%%%%%%%%%%%%%%%%%%%%%%%%%%%%
\label{sec::experiment}
%%%%%%%%%%%%%%%%%%%%%%%%%%%%%%%%%%%%%%%%%%%%%%%%%% 

In this section, we present the results of the simulations and compare them
with experiments of
Ref.~\citenum{Alfred2018}.
The simulation parameters are given in
Table~\ref{tab::params_vals}.

\subsection{Comparison to the short-time transient reflectivity
measurements}
\label{subsec::exp::reflectivity}

\begin{figure}[t!]
\includegraphics[width=9cm]{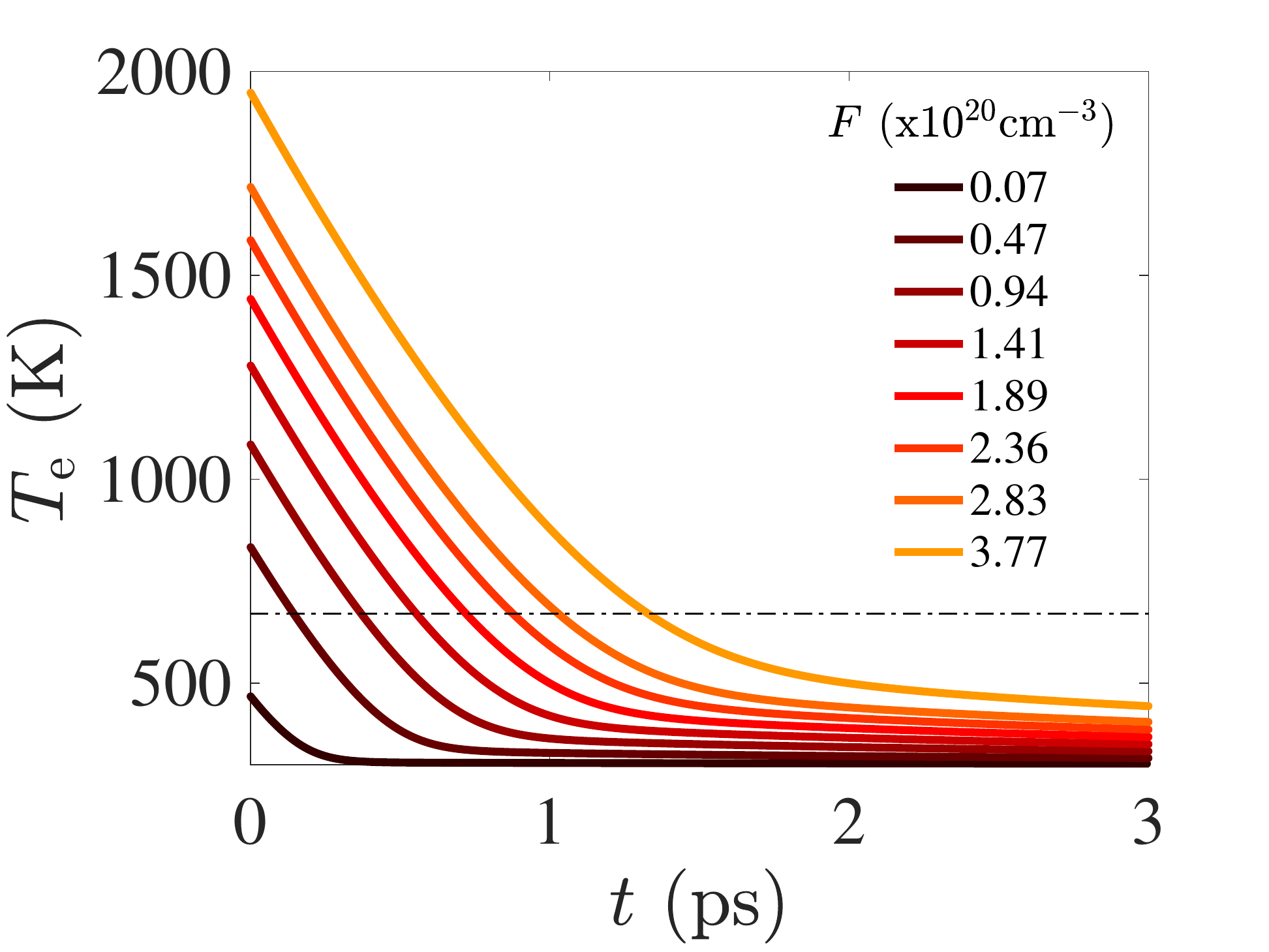}
\caption{ Electronic temperature dynamics for different photoexcitation
densities $F$, as described by the three-temperature model. The in-plot
legend explains the correspondence between $F$ and the curves. Horizontal
dash-dotted line marks the CDW transition temperature 
$T_{\rm c} = 670$\,K.
The crossing of this line with an individual temperature evolution curve 
$T_{\rm e} (t)$
defines time
$\tau_{\rm c} (F)$
introduced by
Eq.~(\ref{eq::tau_tc_def}).
%%%%%%%%%%%%%%%%%%%%%%%%%%%%%%%%%%%%%%%%%%%%%%%%%% 
\label{fig::T_evolution}
%%%%%%%%%%%%%%%%%%%%%%%%%%%%%%%%%%%%%%%%%%%%%%%%%% 
}
\end{figure}

Laser pulse initially excites electronic degrees of freedom, which, in
turn, excite the lattice. Both the electrons and the lattice contribute to
the change in the transient reflectivity signal. We expect that the
monotonically decaying part of the transient reflectivity (measured
transient reflectivity with the oscillating contribution from the amplitude
mode
subtracted~[\!\!~\citenum{Alfred2018}])
tracks the dynamics of the electronic temperature
$T_{\rm e} (t)$.

The computed time dependence
$T_{\rm e} (t)$
is shown in
Fig.~\ref{fig::T_evolution}
for different excitation densities $F$. The crossover between the first
rapid stage to the second slow stage is clearly seen. For longer times
$T_{\rm e}(t)$
approaches some photoexcitation-density-dependent base temperature, which
only slightly exceeds
$T_{\rm env}$
for all excitation densities used.

\begin{figure}[t!]
\includegraphics[width=8.5cm]{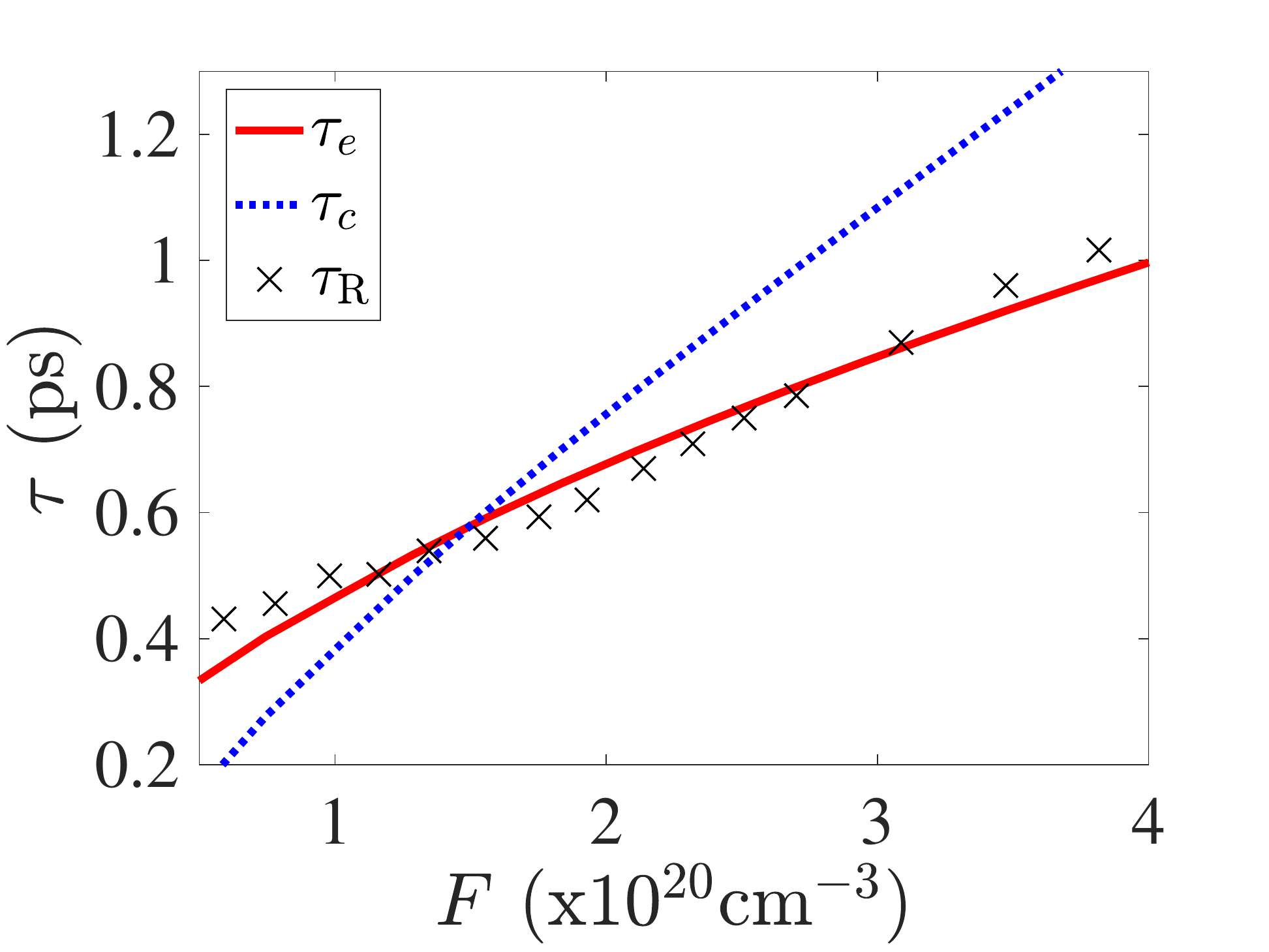}
\caption{Time scales
$\tau_{\rm e}$,
Eq.~(\ref{eq::tau_elec_def}),
and
$\tau_{\rm c}$,
Eq.~(\ref{eq::tau_tc_def}),
as functions of photoexcitation density $F$. Crossed points correspond to
the quasiparticle time
$\tau_{\rm R}$,
extracted from the transient reflectivity
measurements~[\!\!~\citenum{Alfred2018}].
%%%%%%%%%%%%%%%%%%%%%%%%%%%%%%%%%%%%%%%%%%%%%%%%%% 
\label{fig::TIMES}
%%%%%%%%%%%%%%%%%%%%%%%%%%%%%%%%%%%%%%%%%%%%%%%%%% 
}
\end{figure}

To compare the computed temperature evolution with experiment, we introduce
time 
$\tau_{\rm e}$
by condition
\begin{eqnarray}
%%%%%%%%%%%%%%%%%%%%%%%%%%%%%%%%%%%%%%%%%%%%%%%%%%
\label{eq::tau_elec_def}
%%%%%%%%%%%%%%%%%%%%%%%%%%%%%%%%%%%%%%%%%%%%%%%%%%
T_{\rm e}(\tau_{\rm e})
- T_{\rm env}
=
\frac{T_{\rm e} (0) - T_{\rm env}}{e},\ e = 2.718\ldots \ .
\end{eqnarray}
It characterizes the time scale of the electronic temperature cooling down
during the first rapid stage. In
Fig.~\ref{fig::TIMES},
we compare
$\tau_{\rm e}$
with the relaxation time
$\tau_{\rm R}$,
extracted from the transient reflectivity
experiment~[\!\!~\citenum{Alfred2018}].
Given the simplicity of our model, the agreement between the theory and the
experiment is reasonable. It was attained by adjusting the parameter
$G_{\rm eL}$
in
Eq.~(\ref{eq::modified_1tm}),
while other parameters affecting the latter equation were fixed as
described in 
subsection~\ref{subsec::app_choice_params_3temp}.

\subsection{Dynamics of the order parameters}
%%%%%%%%%%%%%%%%%%%%%%%%%%%%%%%%%%%%%%%%%%%%%%%%%%
\label{subsect::Bragg_peak}
%%%%%%%%%%%%%%%%%%%%%%%%%%%%%%%%%%%%%%%%%%%%%%%%%% 

\begin{figure}[t!]
\includegraphics[width=1.0\linewidth]{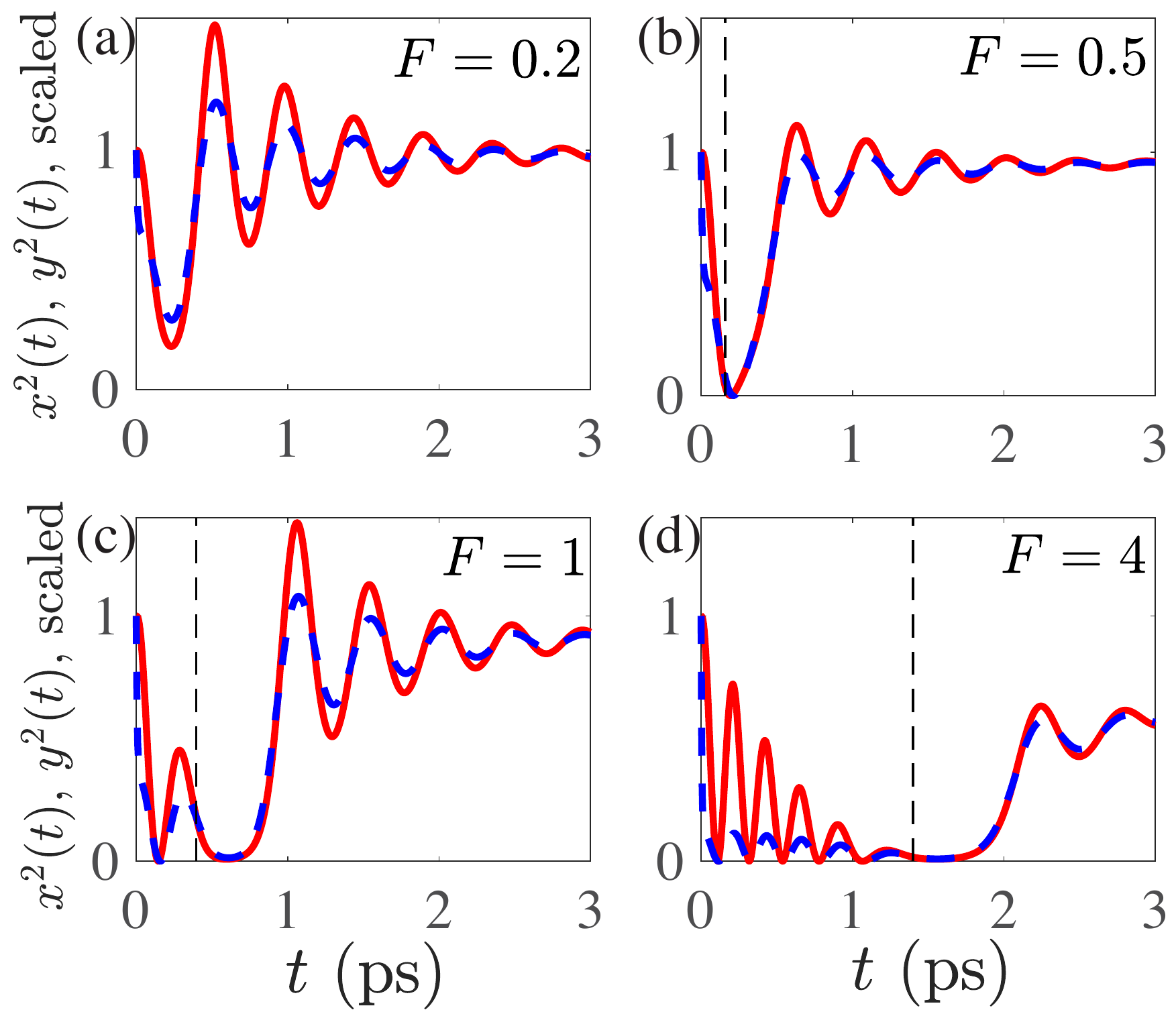}
\caption{Time evolution of the CDW order parameters. Four panels correspond
to different photoexcitation densities $F$ (shown in each panel in units of
10$^{20}$\,cm$^{-3}$).
Dashed blue lines show time-dependence of the electronic CDW order
$x^2(t)$
normalized to its pre-pulse value
$x^2(0^{^{-}})$.
Similarly, solid red lines show the time evolution of the lattice CDW order
represented as
$y^2(t)/y^2(0^{^{-}})$.
Vertical dashed lines mark
$t=\tau_{\rm c} (F)$
defined by
Eq.~(\ref{eq::tau_tc_def}).
%%%%%%%%%%%%%%%%%%%%%%%%%%%%%%%%%%%%%%%%%%%%%%%%%% 
\label{fig::OP_dynamics}
%%%%%%%%%%%%%%%%%%%%%%%%%%%%%%%%%%%%%%%%%%%%%%%%%% 
}
\end{figure}

\subsubsection{Melting of the CDW order}
%%%%%%%%%%%%%%%%%%%%%%%%%%%%%%%%%%%%%%%%%%%%%%%%%%
\label{subsub::melting}
%%%%%%%%%%%%%%%%%%%%%%%%%%%%%%%%%%%%%%%%%%%%%%%%%% 

In
Fig.~\ref{fig::OP_dynamics},
the plots of
$x^2(t)$
and
$y^2(t)$
illustrate the typical dynamics of the electronic and the lattice CDW order
parameters after the arrival of a laser pulse. For low excitation
densities, such as that of
Fig.~\ref{fig::OP_dynamics}a,
the laser pulse does not completely destroys the CDW order -- it only excites
damped AM oscillations around the equilibrium values of the order
parameters. For stronger pulses, as in panels~(b--d), both
$x(t)$
and
$y(t)$
cross zero, which, despite the lack of equilibrium, indicates the proximity
to the melting of the CDW order.

To investigate the onset of the CDW melting, it is useful to look back at
the corresponding temperature evolution
$T_{\rm e} (t)$
shown in
Fig.~\ref{fig::T_evolution},
where we see that
$T_{\rm e} (t)$
stays above
$T_{\rm c}$
for a finite time
$\tau_{\rm c} (F)$.
This time is defined by the condition
\begin{eqnarray}
%%%%%%%%%%%%%%%%%%%%%%%%%%%%%%%%%%%%%%%%%%%%%%%%%%
\label{eq::tau_tc_def}
%%%%%%%%%%%%%%%%%%%%%%%%%%%%%%%%%%%%%%%%%%%%%%%%%%
T_{\rm e}(\tau_{\rm c}) = T_{\rm c},
\end{eqnarray} 
which is indicated in
Fig.~\ref{fig::T_evolution}
by the dash-dotted horizontal line. Once $F$ is larger than a certain
threshold,
$\tau_{\rm c}$
grows monotonically with $F$, as shown in
Fig.~\ref{fig::TIMES}.

In 
Fig.~\ref{fig::OP_dynamics}(b--d),
the time
$t=\tau_{\rm c}$
is indicated by vertical dashed lines. We see that
$\tau_{\rm c}$
indeed gives the correct estimate of the time when the CDW order
ceases to decay and starts to recover. We also note
that for panel~(d), where
$\tau_{\rm c} \gtrsim 1$\,ps,
the order parameters demonstrate multiple passages through zero with
decreasing amplitude. After the oscillations fade, the order parameters
remain suppressed for about 0.5\,ps.

In general, the notion of melting in the course of a non-equilibrium
evolution is not sharply defined. Here we adopt the criterion that the CDW
order undergoes melting when the electronic and the lattice order
parameters do not simply cross zero but rather approach zero in a damped
oscillatory (or non-oscillatory) fashion. From such a perspective,
Fig.~\ref{fig::OP_dynamics}d
represents the melting behavior, while 
Fig.~\ref{fig::OP_dynamics}b
does not, and
Fig.~\ref{fig::OP_dynamics}c
is the border case. On the basis of the above analysis, we conclude that
the critical excitation density 
$F_{\rm c}$,
defined as the lowest border for melting, satisfies the following
inequality
\begin{eqnarray}
%%%%%%%%%%%%%%%%%%%%%%%%%%%%%%%%%%%%%%%%%%%%%%%%%%
\label{eq::F_crit}
%%%%%%%%%%%%%%%%%%%%%%%%%%%%%%%%%%%%%%%%%%%%%%%%%% 
1\times 10^{20}\,{\rm cm}^{-3}
\lesssim F_{\rm c} <
4\times 10^{20}\,{\rm cm}^{-3}.
\end{eqnarray}
Experimentally~[\!\!~\citenum{Alfred2018}]
$F_{\rm c} \sim 2.0 \times 10^{20}\,{\rm cm}^{-3}$,
in agreement with the above constraints.

All plots in
Fig.~\ref{fig::OP_dynamics}
exhibit prominent oscillatory behavior of the order parameters. At lower
photoexcitation densities, such as in
Fig.~\ref{fig::OP_dynamics}a,
the oscillations are clearly related to the
appearance~[\!\!\citenum{higgs_sc2015}]
of the AM observed in the transient reflectivity experiments. However, the
experiment indicates significant reduction of the oscillation amplitude for
$F \gtrsim 2 \times 10^{20}$\,cm$^{-3}$.
This discrepancy can be attributed to our assumption that the order
parameters are homogeneous in space, while in the real system, the spatial
configuration of the order parameters following the melting and the
subsequent re-emergence of the CDW is likely strongly inhomogeneous due to
the appearance of topological defects in the order parameter texture. As a
result of this inhomogeneity, the system has relatively small coherent CDW
domains of varying size with different size-dependent frequencies
$\omega_0$,
which, in turn, leads to the strong dephasing of the oscillations, once the
signal is averaged over the entire sample.

Further analyzing oscillations in
Fig.~\ref{fig::OP_dynamics}d,
we observe that the frequency of transient oscillations for
$t \lesssim 1.5$\,ps
is twice the AM frequency
$\omega_{\rm AM}$.
Such a doubling occurs because of the interplay of two factors: (i)~in
Fig.~\ref{fig::OP_dynamics}
we plot
$x^2(t)$
and
$y^2(t)$
instead of
$x(t)$
and
$y(t)$
and (ii)~the order parameters oscillate near 
$x=y=0$.
The experiments of
Refs.~\citenum{period_doubling2014_prl,period_doubling2014_nature}
indicate that such a frequency doubling may, actually, occur in real
systems.
\begin{figure}[t!]
\includegraphics[width=1.0\linewidth]{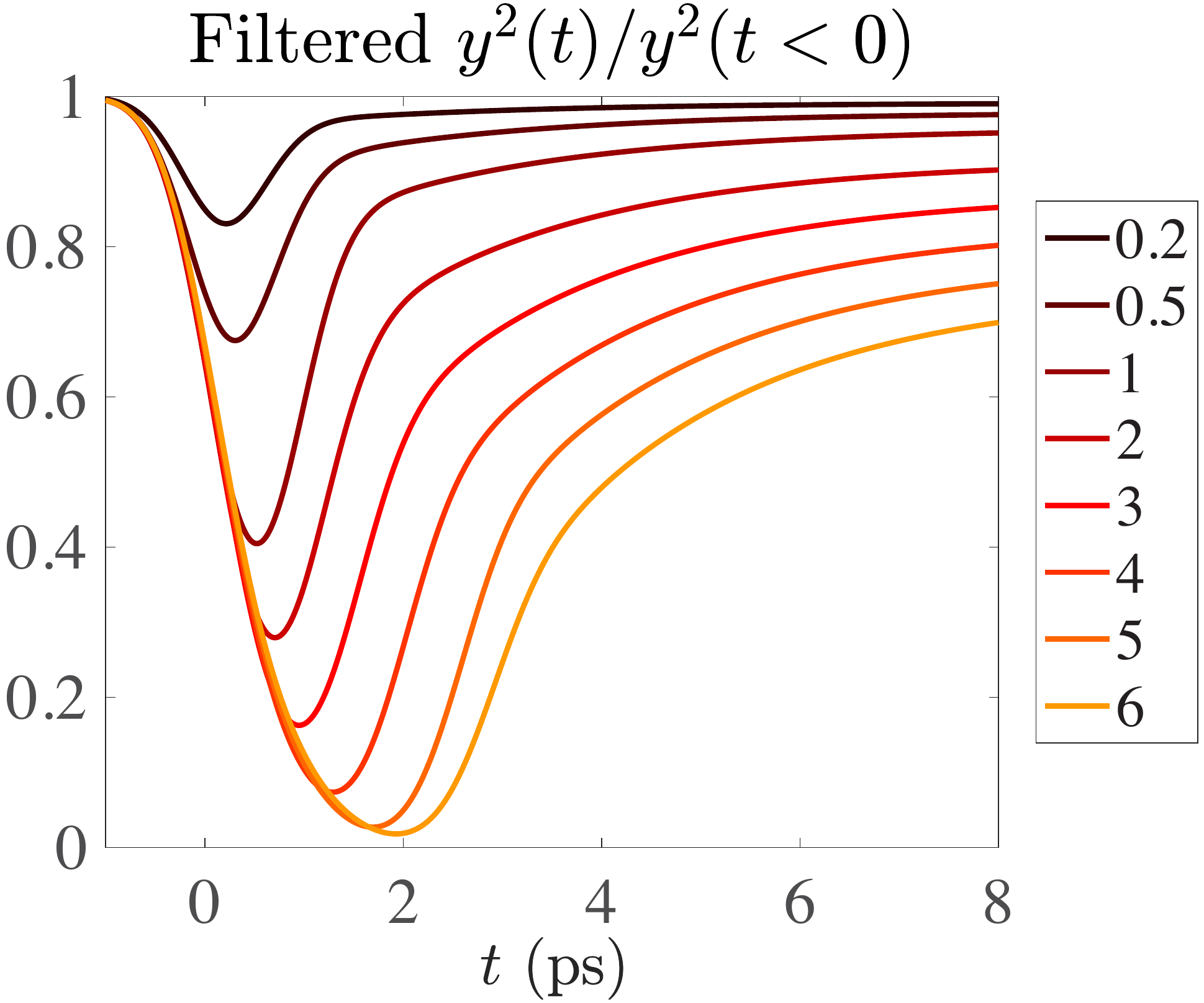}
\caption{Time evolution of the lattice order parameter
$y^2(t)/y^2(t<0)$
-- see
Fig.~\ref{fig::OP_dynamics}
-- filtered with Gaussian function in
Eq.~(\ref{eq::filter})
in order to mimic finite resolution in the UED experiment, for several
values of $F$.
%%%%%%%%%%%%%%%%%%%%%%%%%%%%%%%%%%%%%%%%%%%%%%%%%% 
\label{fig::OP_filtered}
%%%%%%%%%%%%%%%%%%%%%%%%%%%%%%%%%%%%%%%%%%%%%%%%%% 
}
\end{figure}

As for the UED experiments of
Ref.~\citenum{Alfred2018},
they have insufficient time resolution to detect the order parameters
oscillations. To represent the experimental observations, we use the
quantity
$(y^2 * g) (t)$,
where the asterisk denotes time convolution, and 
\begin{eqnarray}
%%%%%%%%%%%%%%%%%%%%%%%%%%%%%%%%%%%%%%%%%%%%%%%%%%
\label{eq::filter}
%%%%%%%%%%%%%%%%%%%%%%%%%%%%%%%%%%%%%%%%%%%%%%%%%% 
g(t) = \frac{1}{\sqrt{2\pi} w}
\exp\!\left(\! -\frac{t^2}{2 w^2}\right)
\end{eqnarray} 
is the Gaussian filter, with parameter $w$ representing the time resolution
of the experiment.
For the UED experiment of
Ref.~\citenum{Alfred2018},
$w=0.42$\,ps.
The results of the convolution are shown in 
Fig.~\ref{fig::OP_filtered}
for different $F$'s. We can see that the oscillations present in all panels
of
Fig.~\ref{fig::OP_dynamics} 
were smeared out by the filter.

\subsubsection{Two kinds of diffraction peaks in UED experiments}
%%%%%%%%%%%%%%%%%%%%%%%%%%%%%%%%%%%%%%%%%%%%%%%%%%
\label{subsub::gen_discuss_UED}
%%%%%%%%%%%%%%%%%%%%%%%%%%%%%%%%%%%%%%%%%%%%%%%%%%

\begin{figure}[t!]
\includegraphics[width=1.0\linewidth]{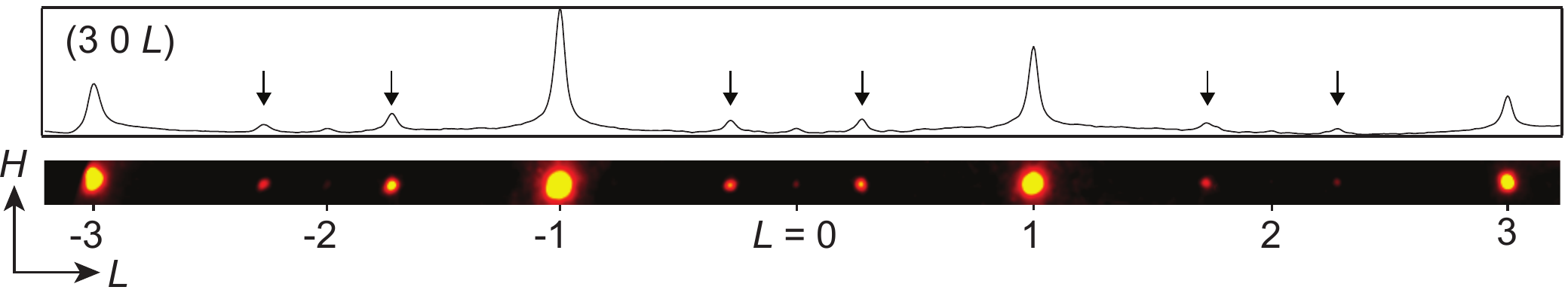}
\caption{Static electron diffraction pattern along (3~0~$L$). The line cut
is obtained by integrating the colored strip along the $H$ direction. The
measured diffraction is a two-dimensional slice in the three-dimensional
reciprocal space. The bright yellow spots are Bragg peaks while arrows mark
the CDW superlattice peaks. Figure reproduced from
Ref.~\citenum{Alfred2018}. 
%%%%%%%%%%%%%%%%%%%%%%%%%%%%%%%%%%%%%%%%%%%%%%%%%% 
\label{fig::UED_exp_spectrum}
%%%%%%%%%%%%%%%%%%%%%%%%%%%%%%%%%%%%%%%%%%%%%%%%%% 
}
\end{figure}

The UED
experiments~[\!\!~\citenum{Alfred2018}]
observe two kinds of diffraction peaks associated either with the
underlying crystal structure of
LaTe$_3$
or with the CDW order, see
Fig.~\ref{fig::UED_exp_spectrum}.
The measurements of
Ref.~\citenum{Alfred2018}
were done in the higher-order Brillouin zones, which implies that the
measured intensities of the CDW peaks are
determined~[\!\!~\citenum{Zheng2005}]
by the lattice CDW order $y$. Fundamentally, the integrated intensity of a
CDW peak is proportional to
$y^2$.
Therefore, at first sight, the direct way to test our modeling is to
compare the calculated
$y^2(t)$
with the time evolution of the integrated CDW peak intensity measured in
the UED experiment. 

However, the problem here is
that the experiment~\citenum{Alfred2018} measured diffraction intensities only for a two-
dimensional slice $(k_x, k_z)$ of a three-dimensional reciprocal space $(k_x, k_y, k_z)$, see Fig. 7. In other words, Ref.~\citenum{Alfred2018} does not contain  direct experimental information about the three-dimensional integrated intensity of the CDW peaks.

At the same time, since, in the
experiment~[\!\!~\citenum{Alfred2018}],
crystal Bragg peaks are resolution-limited, the two-dimensional integrals
over the measured sections of these peaks are proportional to the
respective integrals over the full three-dimensional reciprocal space. One
can now take into account the sum rule implying that the emergence of the
CDW order leads to the intensity transfer from the Bragg peaks to the CDW
peaks, see
Appendix~\ref{app::bragg_sum_rule}.
Therefore, in our case, the most direct way to extract the value of 
$y^2(t)$
from experiment is to examine the integrated intensity by the Bragg peaks:
the suppression of the CDW order leads to increase of the Bragg peak
intensity. This relation is quantified in the next subsection. 

\subsubsection{Time evolution of the Bragg peaks for the underlying crystal
lattice from UED experiments}
%%%%%%%%%%%%%%%%%%%%%%%%%%%%%%%%%%%%%%%%%%%%%%%%%%
\label{subsub::bragg_evolution}
%%%%%%%%%%%%%%%%%%%%%%%%%%%%%%%%%%%%%%%%%%%%%%%%%%

For the integrated Bragg peak intensity $I$ in the presence of the CDW
order, we use the expression obtained in
Ref.~\citenum{Overhauser1971Observability}:
\begin{eqnarray}
%%%%%%%%%%%%%%%%%%%%%%%%%%%%%%%%%%%%%%%%%%%%%%%%%% 
\label{eq::I_bragg}
%%%%%%%%%%%%%%%%%%%%%%%%%%%%%%%%%%%%%%%%%%%%%%%%%% 
I \propto [J_0(p y)]^2 e^{-2 W},
\end{eqnarray}
where 
$J_0$
is the zeroth order Bessel function, and $p$ is a constant. Parameter $W$
accounts for the Debye-Waller suppression of the intensity due to thermal
fluctuations of the phonons. 

We expect that
$W \propto T_{\rm L1}$,
and, therefore, write
\begin{eqnarray}
%%%%%%%%%%%%%%%%%%%%%%%%%%%%%%%%%%%%%%%%%%%%%%%%%%
\label{eq::BP0}
%%%%%%%%%%%%%%%%%%%%%%%%%%%%%%%%%%%%%%%%%%%%%%%%%%
\frac{I(t)}{I(0)}
&=&
\frac{J_0 (py(t))}{J_0 (py(0))} e^{-2 [W(t)-W(0)]}
\approx \notag{}
\\
&\approx& 1 - P [y^2(t) - y^2_{\rm eq}] - S [T_{\rm L1}(t) - T_{\rm env}],
\end{eqnarray}
where
$P = p^2/2$,
and $S$ is a constant.

As in
subsection~\ref{subsub::melting},
to account for the finite experimental temporal resolution, we convolute
the rhs of
Eq.~(\ref{eq::BP0})
with the Gaussian
filter~(\ref{eq::filter}).
The final expression used to mimic the actual Bragg peaks dynamics reads
\begin{eqnarray}
\frac{I(t)}{I(t_0)}
&\approx&
\left\{ 
		1 - P [y^2(t-t_0) - y^2_{\rm eq}] 
		\right.
\nonumber \\
		&&\left.
- S [T_{\rm L1}(t-t_0) - T_{\rm env}]
		\right\} * g.
%%%%%%%%%%%%%%%%%%%%%%%%%%%%%%%%%%%%%%%%%%%%%%%%%%
\label{eq::BP}
%%%%%%%%%%%%%%%%%%%%%%%%%%%%%%%%%%%%%%%%%%%%%%%%%%
\end{eqnarray}
Here 
$t_0$
is an adjustable parameter shifting the origin of the time axis. This shift
is another consequence of the limited time resolution of the
experiment~[\!\citenum{timeshift_note}].

\begin{figure}[t!]
\includegraphics[width=1\linewidth]{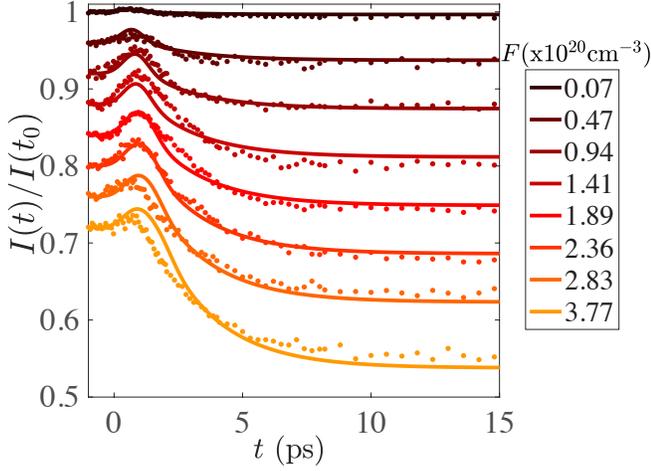}
\caption{Time evolution of the Bragg peaks intensity
$I(t)$
for different excitation densities. Solid lines are obtained from
Eq.~(\ref{eq::BP});
dots represent the experimental
data~[\!\!~\citenum{Alfred2018}].
The lines and the data points for different excitation densities are
vertically displaced for clarity.
%%%%%%%%%%%%%%%%%%%%%%%%%%%%%%%%%%%%%%%%%%%%%%%%%% 
\label{fig::Adiabatic_BP}
%%%%%%%%%%%%%%%%%%%%%%%%%%%%%%%%%%%%%%%%%%%%%%%%%% 
}
\end{figure}

Function
$I(t)$,
numerically evaluated with the help of
Eq.~(\ref{eq::BP}),
together with the UED data points, are plotted in
Fig.~\ref{fig::Adiabatic_BP}.
All eight plots in the latter figure were obtained by fitting the
experimental points using four adjustable parameters:
$\kappa = 0.2$,
$P=0.1$,
$S =3\times 10^{-3}$,
and
$t_0 =0.43$\,ps.
The above value of parameter $\kappa$ was used for all simulations
presented in this paper.

Overall, the agreement between the fits and the experiment in
Fig.~\ref{fig::Adiabatic_BP}
is rather good. For higher excitation densities $F$, the small discrepancy
might be due to the fact that 
approximation~(\ref{eq::electr_heat_cap})
for the electronic heat capacity is less accurate at higher temperatures.

\subsubsection{Time evolution of the CDW peak from UED experiments}
%%%%%%%%%%%%%%%%%%%%%%%%%%%%%%%%%%%%%%%%%%%%%%%%%%
\label{subsub::cdw_evolution}
%%%%%%%%%%%%%%%%%%%%%%%%%%%%%%%%%%%%%%%%%%%%%%%%%% 

\begin{figure}
\begin{minipage}[h]{1\linewidth}
\center{\includegraphics[width=1\linewidth]{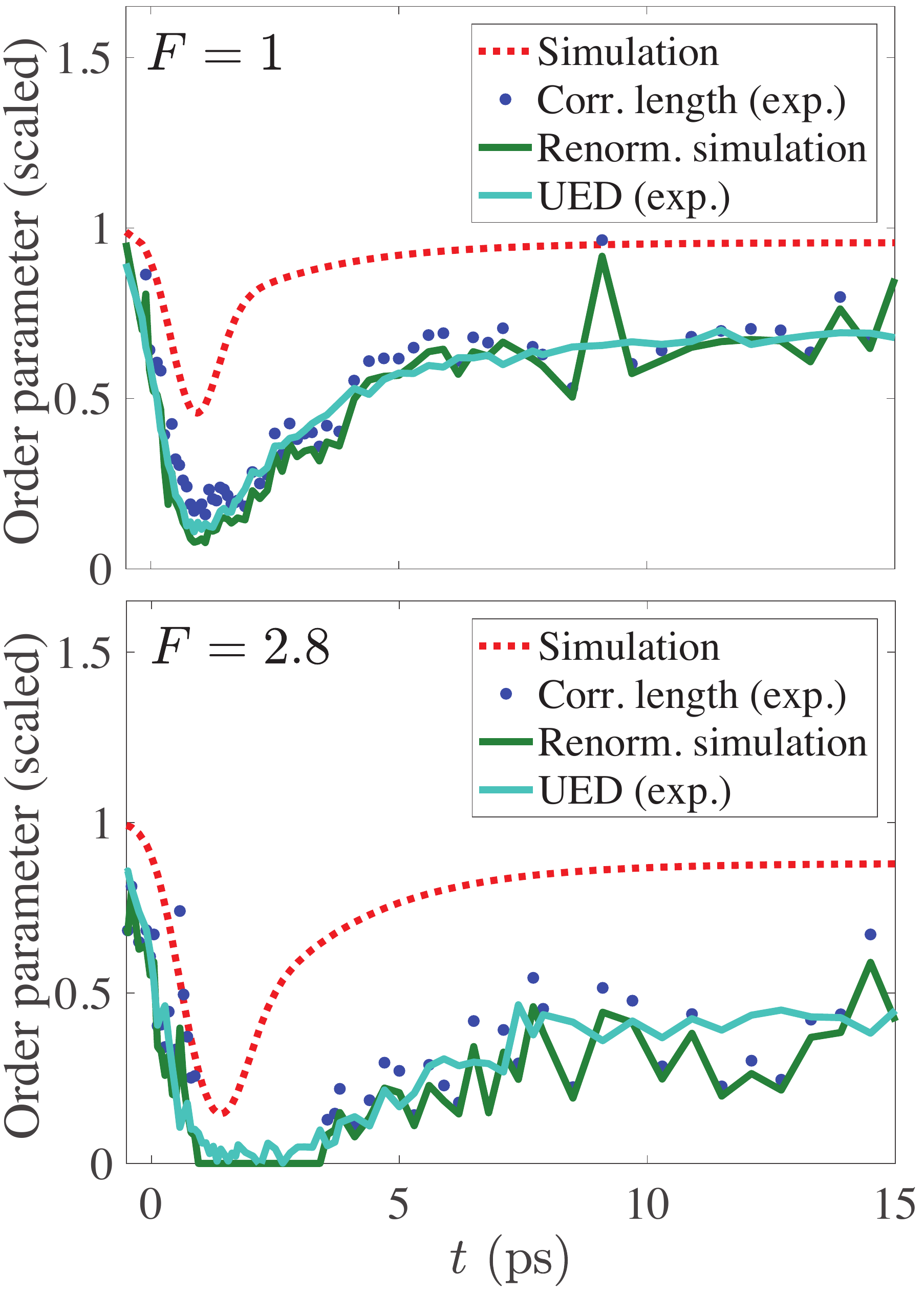}} \\
\end{minipage}
\caption{Comparison of the simulated dynamics with the UED data for the CDW
peak. Two panels correspond to two different photoexcitation densities:
$F=9.4\times 10^{19}$\,cm$^{-3}$
(top panel) and 
$F=2.8\times 10^{20}$\,cm$^{-3}$
(bottom). Quantity
$(y^2*g)(t)$,
representing simulated filtered dynamics of the order parameter (see
Fig.~\ref{fig::OP_filtered}),
is shown by dashed (red) curve. Experimentally
obtained~[\!\!~\citenum{Alfred2018}]
partially integrated UED intensity
$I^{\rm CDW}_{\rm 2D}$,
Eq.~(\ref{eq::I_xy_def}),
is shown by solid (cyan) curves. The data 
points~[\!\!~\citenum{Alfred2018}]
for the CDW correlation length
$\xi$ are shown as (blue) dots. For larger
$F$ (bottom panel), the CDW peak disappears for
1\,ps\,$\lesssim t\lesssim$3.5\,ps,
consequently, the data points for
$\xi$ are absent in this interval. To
account for the theoretically unknown phase dynamics, we multiply
$(y^2\!*g)$
by
$\xi$ [as in
Eq.~(\ref{eqn::CDW_int_peak})].
The resulting dependence is shown by solid (green) line. For both
$F$'s,
the agreement between
$I^{\rm CDW}_{\rm 2D}$
and
$(y^2\!*g)\, \xi$
is quite notable.
%%%%%%%%%%%%%%%%%%%%%%%%%%%%%%%%%%%%%%%%%%%%%%%%%% 
\label{fig::renorm_ord_params}
%%%%%%%%%%%%%%%%%%%%%%%%%%%%%%%%%%%%%%%%%%%%%%%%%%
}
\end{figure}

We now turn to the discussion of the CDW peak.
%which, as explained in
%subsection~\ref{subsub::gen_discuss_UED},
%contains information about the long-range lattice order. It tracks both the
%amplitude and the phase of the CDW, cf.
%Eq.~(\ref{eqn::App_CDW_peak}).
As anticipated in
subsection~\ref{subsub::gen_discuss_UED},
a straightforward attempt to approximate the experimentally measured
$(k_x,k_z)$-integrated
intensity of a CDW peak by a theoretically computed 
$y^2(t)$
[with the appropriate convolution and time shift, as in
Eq.~(\ref{eq::BP})]
reveals large discrepancy. Indeed, as one can see from
Fig.~\ref{fig::renorm_ord_params},
the calculated curves (red dashed lines) lie significantly higher than the
experimental UED data. Below, we test the proposition formulated in
Ref.~\citenum{Alfred2018}
that the deviations between the data and the simulations originate from the
fact that the UED measurements access only a two-dimensional
$(k_x, k_z)$-slice
of the three-dimensional 
$(k_x, k_y, k_z)$-space.

We assume that the intensity of the CDW peak in the reciprocal space can be
reasonably approximated by a factorized function
\begin{eqnarray}
%%%%%%%%%%%%%%%%%%%%%%%%%%%%%%%%%%%%%%%%%%%%%%%%%%
\label{eq::CDW_lorentz}
%%%%%%%%%%%%%%%%%%%%%%%%%%%%%%%%%%%%%%%%%%%%%%%%%% 
{\cal G}^{\rm CDW}({\bf k})
\propto
y^2 s_x ( k_x) s_y( k_y) s_z( k_z)
\end{eqnarray}
where 
$s_{\sigma } (k_\sigma )$
are the peak shape functions for the respective $k$-space directions, with
index $\sigma$ taking values $x$, $y$, or $z$. Following the convention of
Ref.~\citenum{Alfred2018},
axes $x$ and $z$ denote the directions parallel to Te$_2$ planes, while
axis $y$ is perpendicular to these planes.
(The notations $x$ and $y$ for the spatial axes appear only as subscripts,
and are not to be confused with the variables $x$ and $y$ defined by 
Eq.~(\ref{dimensionless})
that represent the electronic and the lattice order parameters.) Functions 
$s_\sigma (k_\sigma)$
are normalized by the condition
\begin{eqnarray}
\int dk s_\sigma (k) = 1.
\end{eqnarray} 
We also assume that these functions are non-negative and bell-shaped. (In
experiment, these functions are fitted by Lorentzians due to the intrinsic
profile of the electronic beam.) The two-dimensional
integral of the CDW peak reported in
Ref.~\citenum{Alfred2018}
can be written as
\begin{eqnarray}
%%%%%%%%%%%%%%%%%%%%%%%%%%%%%%%%%%%%%%%%%%%%%%%%%%
\label{eq::I_xy_def}
%%%%%%%%%%%%%%%%%%%%%%%%%%%%%%%%%%%%%%%%%%%%%%%%%% 
I^{\rm CDW}_{\rm 2D}
\equiv
\int d k_x d k_z {\cal G}^{\rm CDW}({\bf k}) |_{k_y = 0}
=
y^2 s_y(0).
\end{eqnarray}
Now we observe that, for common bell-shaped functions, such as Lorentzian
or Gaussian, 
$s_\sigma(0) \propto \xi_\sigma$,
where
$\xi_\sigma$
is the real-space correlation length in the respective direction. Using
this observation and adding explicit time dependencies of the parameters
involved, we arrive at the expression
\begin{eqnarray}
I^{\rm CDW}_{\rm 2D}(t) \propto y^2(t) \xi_y(t).
\end{eqnarray} 

In this work, we assume that the order parameter is homogeneous is space.
Thus, we cannot obtain theoretically 
$\xi_y (t)$.
However, we can estimate it on the basis of the assumption that all
correlation lengths are determined by the same mechanism and hence are
proportional to each other, i.e.
\begin{eqnarray}
%%%%%%%%%%%%%%%%%%%%%%%%%%%%%%%%%%%%%%%%%%%%%%%%%%
\label{eq::xi_assumption}
%%%%%%%%%%%%%%%%%%%%%%%%%%%%%%%%%%%%%%%%%%%%%%%%%% 
\xi_y (t) \propto \xi_{x,z}(t) \approx \xi_{\rm exp} (t),
\end{eqnarray} 
where
$\xi_{\rm exp} (t)$
is the experimentally measured correlation length in the $x$- and
$z$-directions, obtained as the inverse of the FWHM of the CDW peak after
instrumental resolution is taken into consideration, see
Eq.~(S4) 
of the Supplementary Information to
Ref.~\citenum{Alfred2018}.

Finally, taking into account the finite experimental time resolution,
as in 
Eq.~(\ref{eq::BP}),
the measured integrated intensity can be approximated as
\mbox{
$I^{\rm CDW}_{\rm 2D, exp}(t) = (y^2\! * g) (t-t_0)\,\, \xi_{\rm exp}(t)$}.
To facilitate the comparison with experiments, we re-express this relation
in the following manner
\begin{eqnarray}
%%%%%%%%%%%%%%%%%%%%%%%%%%%%%%%%%%%%%%%%%%%%%%%%%% 
\label{eqn::CDW_int_peak}
%%%%%%%%%%%%%%%%%%%%%%%%%%%%%%%%%%%%%%%%%%%%%%%%%% 
\frac{ I^{\rm CDW}_{\rm 2D, exp}(t)}{ I^{\rm CDW}_{\rm 2D, exp}(0^{^{-}})}
=
\frac{(y^2\! * g) (t-t_0)\,\, \xi_{\rm exp}(t)}
	{(y^2\! * g) (0^{^{-}})\,\, \xi_{\rm exp}(0^{^{-}})},
\end{eqnarray}
where the argument 
$(0^{^{-}})$
implies the pre-pulse values of the respective parameters. 

In
Fig.~\ref{fig::renorm_ord_params},
we test the
relation~({\ref{eqn::CDW_int_peak}})
by substituting there the theoretically calculated
$y^2(t)$.
The agreement between the direct measurement and the prediction of
Eq.~(\ref{eqn::CDW_int_peak}) 
is rather encouraging. This is another consistency check of our modeling.

\section{Overview of time scales}
%%%%%%%%%%%%%%%%%%%%%%%%%%%%%%%%%%%%%%%%%%%%%%%%%%
\label{sec::time_scales}
%%%%%%%%%%%%%%%%%%%%%%%%%%%%%%%%%%%%%%%%%%%%%%%%%%

\begin{figure}
\begin{minipage}[h]{1\linewidth}
%\center{\includegraphics[width=1.1\linewidth]{arrow}} \\
\center{\includegraphics[width=1\linewidth]{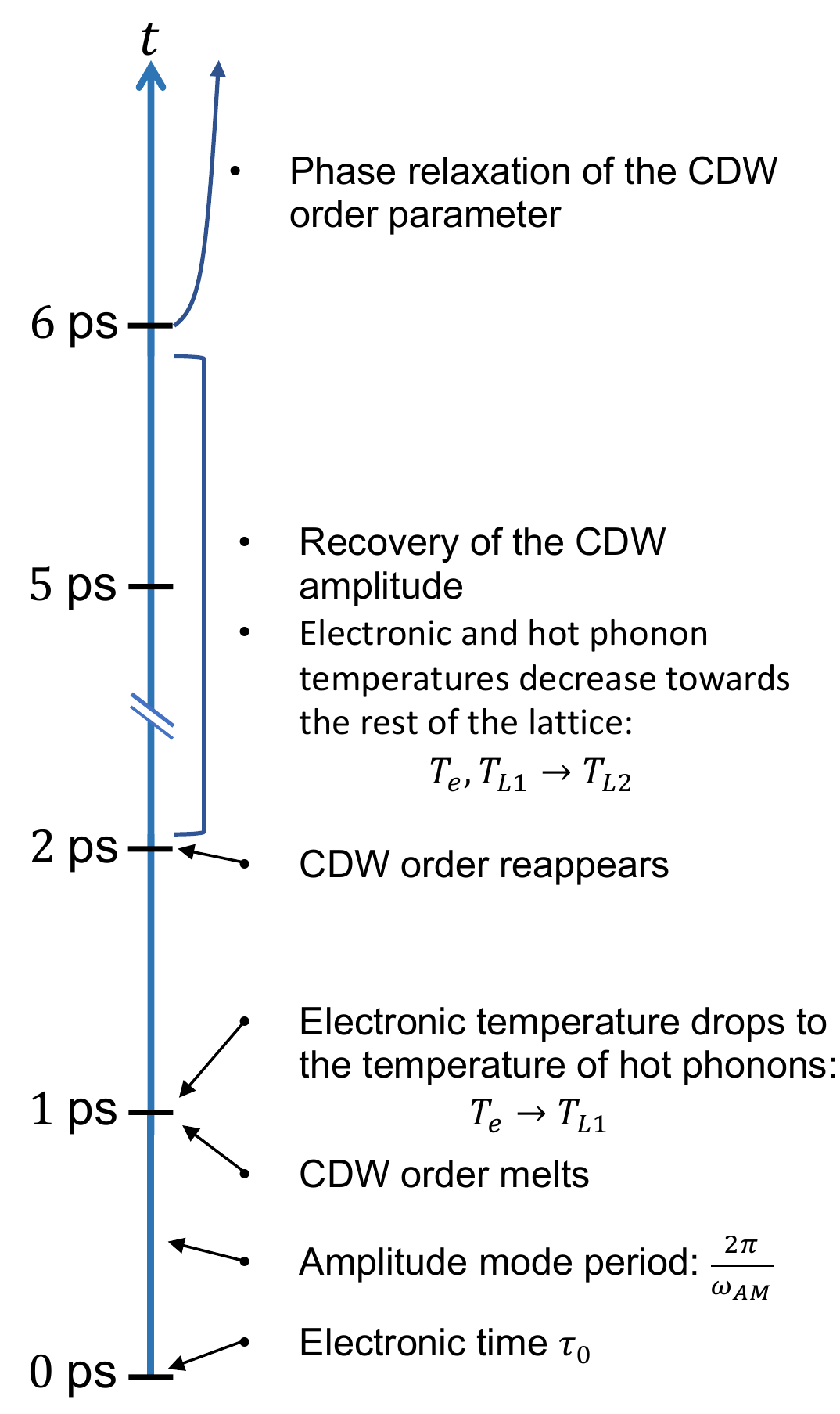}} \\
\end{minipage}
\caption{Sketch of the stages of nonequilibrium time evolution in the
course of a photoexcited phase transition in LaTe$_3$. The characteristic
times indicated in the figure are the estimates made in
Section~\ref{sec::time_scales} for $F=3 \times 10^{20}$~cm$^{-3}$.
%%%%%%%%%%%%%%%%%%%%%%%%%%%%%%%%%%%%%%%%%%%%%%%%%% 
\label{fig::time_scales}
%%%%%%%%%%%%%%%%%%%%%%%%%%%%%%%%%%%%%%%%%%%%%%%%%%
}
\end{figure}

Here we would like to bring together various aspects of our simulations by
attaching concrete time scales to the general non-equilibrium scenario
described in
Section~\ref{preliminary}.
In order to be specific, we choose the case of the photoexcitation density
$F=3$,
which is above the transient-melting threshold
$F_c=2$
determined in
Sec.~\ref{subsub::melting}. (Here and below, the units for $F$ are $10^{20}\text{cm}^{-3}$.)
Different stages of the non-equilibrium evolution together with the
relevant characteristic times are summarized in
Fig.~\ref{fig::time_scales}.

The fastest time appearing in our description is
$\tau_0 = 20$\,fs.
It enters
Eq.~(\ref{eq::TDGL1}),
and characterizes the relaxation time of the electronic
density~[\!\citenum{thermalize_note}].
Among time scales of the lattice dynamics, the shortest one is the
period of the amplitude mode
$2 \pi/\omega_{\rm AM} \sim 0.5$\,ps.
Next is the relaxation time of the CDW phonon mode 
$1/\gamma_y \omega_0 \sim 1$~ps.
Another relevant time having the value of approximately 1~ps is 
$\tau_{\text{e}}$, which
describes the convergence of the electronic temperature
$T_{\rm e}$
and the temperature of the hot phonons subsystem
$T_{\rm L2}$ (see Fig.~\ref{fig::TIMES}).

For the photoexcitation density
$F=3$,
the electronic temperature
$T_{\rm e}$
initially jumps to a value significantly higher than
$T_{\text{c}}$
and then  remains above
$T_{\rm c}$
for a time longer than 1~ps. This leaves enough time for  the lattice CDW
order to relax to zero, which implies complete melting. By the time of about 2~ps, two changes occur:  (i) Electronic temperature
$T_{\rm e}$
drops below
$T_{\rm c}$,
which leads to the reappearance of both the electronic and the lattice CDW
orders. (ii) Simultaneously, the character of the temperature relaxation
changes ---  the electronic temperature
$T_{\rm e}$
and the hot phonons temperature
$T_{\rm L2}$,
after having approached each other, start decreasing towards the
temperature of the rest of the lattice
$T_{\rm L1}$
with characteristic time
$\tau_{\rm DW} \approx 2$\,ps.
The recovery of the electronic and the lattice CDW amplitudes becomes
eventually completed by the time of about 6~ps. (Due to small overall
heating, the CDW amplitude recovers to a value, which is  slightly smaller
than the pre-pulse one.)
 
Finally, as indicated in
Fig.~\ref{fig::time_scales}, 
the longest time scale is supposed to be the one associated with the relaxation
of the phase of the CDW order parameter. Since our formalism deals exclusively
with the dynamics of the order parameter amplitude, and not the phase,
these slow relaxation processes are not covered by our simulations.

\section{Discussion}
%%%%%%%%%%%%%%%%%%%%%%%%%%%%%%%%%%%%%%%%%%%%%%%%%%
\label{sec::discussion}
%%%%%%%%%%%%%%%%%%%%%%%%%%%%%%%%%%%%%%%%%%%%%%%%%% 

In this paper, we have shown that the combination of the
three temperature model and the TDGL equations with two order parameters
constitutes a quantitatively adequate description of the amplitude
response of CDW materials to a strong femtosecond laser pulse. The
successful description of the time evolution of the crystal Bragg peaks
intensities shown in
Fig.~\ref{fig::Adiabatic_BP}
is the testament of the model's predictive power. Indeed, we were able to
simulate a family of eight different time-dependent intensities
corresponding to different photoexcitation densities using only one physically important parameter $\kappa$ and three  ``technical'' parameters  $P$, $S$, and
$t_0$.
The values of other parameters required by the model were obtained from
independent experiments on CDWs in
LaTe$_3$
and in other rare-earth  tritellurides. We thus demonstrated how to use the 
available experimental knowledge to systematically extract the parameters for 
far-from-equilibrium CDW simulations. 

We further note that our simulations predict that electronic and lattice
modulations are supposed to exhibit an oscillatory dynamics in the vicinity
of the CDW melting transition with a frequency which is two times larger
than that of the weakly perturbed CDW (see
Fig.~\ref{fig::OP_dynamics}
and the discussion in
Sec.~\ref{subsub::melting}).
This dynamics would not be directly visible in experiments of
Ref.~\citenum{Alfred2018}
due to finite time resolution but, otherwise, is consistent with the
experiments~[\!\!~\citenum{period_doubling2014_prl,
period_doubling2014_nature}]
with the blue bronze
K$_{0.3}$MoO$_3$
and a perovskite-type manganite
Pr$_{0.5}$Ca$_{0.5}$MnO$_3$.
In a related development, the application of the present simulation
framework in
Ref.~\citenum{AlfredPavelPRL} 
pointed at the existence of a dynamical slowing down regime near the CDW
melting transition, where the order parameters become ``caught'' near the
metastable maximum of the free
energy~(\ref{eq::landau_func}).

In terms of advancing the general knowledge about the far-from-equilibrium
CDW dynamics, our simulations shed light on the transient CDW response that
is difficult to access experimentally. In particular, the necessity to use
two different lattice temperatures --- one for hot phonons and the other
for the rest of the lattice ---  confirms previous
conjectures~[\!\!\citenum{tbte3_phonon_maschek_exp2015,dyte3_phonon_maschek_exp2018,
mansart_3temp_pnas2012,h_nbs2_electr_phon_coupl_anisotr2012,
nonuniform_electron_phon_coupl_nbse2,review_ultrafast2017}]
that the
energy transfer from the photo-excited electrons to the phonon bath occurs
unevenly among different phonon modes. Overall, the developed theoretical framework should be applicable to other CDW materials
and other experimental settings, such as three-pulse experiments of
Ref.~\citenum{yusupov_nat_phys2010}.

Although our treatment only deals with the amplitudes of the order
parameters, an important outcome of this work is that, for experimental
quantities affected by both the amplitude and the phase relaxation of the
CDW order, the lack of the theoretical information about the phase dynamics
may be compensated, at least partially, using phenomenology-based approach.
An example here is the comparison between the measured 2D-integrated UED
intensity of the CDW peak presented in
Fig.~\ref{fig::renorm_ord_params} and the computed intensity.
We interpret the difference between the two as being caused by the phase
relaxation of the CDW order, which is slower than the amplitude relaxation
due to the possible presence of topological defects --- in agreement with
the analysis of
Ref.~\citenum{Alfred2018}.
The consistency of this interpretation is further demonstrated in
Fig.~\ref{fig::renorm_ord_params}
by correcting the amplitude-dependent intensity with a factor determined
from the experimental knowledge of the peak width, which is, in turn,
determined by the phase fluctuations of the order parameter. 

%\subsection{Limitations of the proposed framework}
%%%%%%%%%%%%%%%%%%%%%%%%%%%%%%%%%%%%%%%%%%%%%%%%%% 
%\label{limitations}
%%%%%%%%%%%%%%%%%%%%%%%%%%%%%%%%%%%%%%%%%%%%%%%%%% 

As far as the weak points of our modeling are concerned, one of them is its 
mean-field character. In reality, many CDW systems, including
LaTe$_3$,
demonstrate pronounced non-mean-field  properties near the phase
transition~[\!\!~\citenum{girault_cdw_crit_expon_exp1989,
ru_3Te_chem_pressure2008,hoesch_cdw_crit_expon_exp2009}].
Fluctuations relative to the mean-field state can be split into two groups:
those of the order parameters amplitudes, and those of the phase. The fluctuations
of the amplitudes are less worrisome: we expect them to be effectively included
into renormalization of the model's coefficients by replacing some (unknown) ``bare" values with the ``effective"
(observable) values. At the same time, phase fluctuations of the CDW order
parameter are of greater concern. They remain largely unaccounted in the
present work, which focuses on the CDW amplitudes.

Another limitation of proposed approach is related to the fact that the
three-temperature model greatly oversimplifies the kinetic processes
in the studied system. However, our success in reproducing the experiments
in general and the transient reflectivity experiments in particular
suggests that this approximation captures essential physics of the system.

\section{Conclusions}
%%%%%%%%%%%%%%%%%%%%%%%%%%%%%%%%%%%%%%%%%%%%%%%%%% 
\label{sec::conclusions}
%%%%%%%%%%%%%%%%%%%%%%%%%%%%%%%%%%%%%%%%%%%%%%%%%% 
We developed a theoretical framework to describe the dynamics of the CDW
amplitude after an intense laser pulse. The framework consists of
(i)~the time-dependent Ginzburg-Landau equations for the electron and
lattice CDW amplitudes and (ii)~the three-temperature model. We tested
the resulting description by comparing the simulations with the available 
experimental data. The agreement is good, suggesting that the proposed 
framework can be applied to a broader class of non-equilibrium settings.

\section{Acknowledgements}
We thank T.~Rohwer, C.~Lee, E.~J.~Sie, E.~Baldini, B.~Freelon, and H.~Zhou
for the help in building and acquiring data from the ultra-fast electron
diffraction and time-resolved ARPES setups. We acknowledge high-quality
samples prepared by J.~Straquadine, P.~Walmsley, I.~R.~Fisher, Y.-Q.~Bie,
and P.~Jarillo-Herrero. 
All authors acknowledge the support of the 
Skoltech NGP Program (Skoltech-MIT joint project) (theory).
N.G., A.Z., and A.K. also acknowledge the support from
the Gordon and Betty Moore Foundations EPiQS Initiative grant GBMF4540
(data analysis).

\appendix

\section{Bragg and CDW diffraction peaks}
%%%%%%%%%%%%%%%%%%%%%%%%%%%%%%%%%%%%%%%%%%%%%%%%%%
\label{app::bragg_sum_rule}
%%%%%%%%%%%%%%%%%%%%%%%%%%%%%%%%%%%%%%%%%%%%%%%%%%

The purpose of this Appendix is to illustrate how the presence of the CDW
modifies electron diffraction peaks. We then discuss the role of
fluctuations of the phase of the order parameter. The discussion here is a
simplified version of a more general treatment of
Ref.~\citenum{Overhauser1971Observability}.

We express modulation of the lattice site positions as
\begin{eqnarray}
{\bf r}_n \rightarrow {\bf r}_n + {\bf u} \cos({\bf Q}\cdot{\bf r}_n),
\end{eqnarray}
where
$\bf u$
and
$\bf Q$
are the amplitude and the wave vector of the modulation;
${\bf r}_n$
are high-symmetry lattice points. Assuming that the amplitude
$|{\bf u}|$
is much smaller than the lattice spacing, we write the density as
\begin{eqnarray}
\rho ( {\bf r} )
&=&
\sum_{ {\bf r}_n } 
	\delta \left[ 
		{\bf r} - {\bf r}_n - {\bf u} \cos({\bf Q}\cdot{\bf r}_n) 
	\right]
\approx
\notag{}
\\
&\approx&
\rho_0 ( {\bf r} )
-
\sum_{ {\bf r}_n } 
	({\bf u}\cdot \nabla_{\bf r} )\delta ( {\bf r} - {\bf r}_n )
	\cos({\bf Q}\cdot{\bf r}_n)  + \notag{}
\\
&+&
\frac{1}{2}
\sum_{ {\bf r}_n } 
	({\bf u}\cdot \nabla_{\bf r} )^2\delta ( {\bf r} - {\bf r}_n )
	\cos^2({\bf Q}\cdot{\bf r}_n) + \dots
\end{eqnarray}
where
$\rho_0 ( {\bf r} )
=
\sum_{ {\bf r}_n } \delta ( {\bf r} - {\bf r}_n)$
corresponds to the density of unmodulated lattice. By performing the
Fourier transformation we obtain:
\begin{widetext}
\begin{flalign}
\rho_{\bf k}
=
\rho_{0 {\bf k}}
&-\int d^3 {\bf r} e^{ - i {\bf k}\cdot{\bf r} }
	\sum_{ {\bf r}_n } 
		({\bf u}\cdot \nabla_{\bf r} )
		\delta ( {\bf r} - {\bf r}_n)\cos({\bf Q}\cdot{\bf r}_n)
		 +
\nonumber\\
&+\frac{1}{2}
\int d^3 {\bf r} e^{ - i {\bf k}\cdot{\bf r} }
	\sum_{ {\bf r}_n } 
		({\bf u}\cdot \nabla_{\bf r} )^2
		\delta ( {\bf r} - {\bf r}_n )
		\cos^2({\bf Q}\cdot{\bf r}_n) + \dots
\end{flalign}
\end{widetext}
where
$\rho_{0\bf k} = F_{\bf k} \sum_{\bf b} \delta_{{\bf k}, {\bf b} }$
is a sum of sharp peaks located at reciprocal wave vectors
$\bf b$
of the underlying crystal lattice. Here 
$F_{\bf k}$
is the lattice form-factor. Integrating by parts, we obtain
\begin{eqnarray}
%%%%%%%%%%%%%%%%%%%%%%%%%%%%%%%%%%%%%%%%%%%%%%%%%% 
\label{eq::rho_k}
%%%%%%%%%%%%%%%%%%%%%%%%%%%%%%%%%%%%%%%%%%%%%%%%%% 
\rho_{\bf k} &&= \rho_{0 {\bf k}} -
i ({\bf u}\cdot {\bf k} ) 
	\sum_{ {\bf r}_n } 
		e^{ - i {\bf k}\cdot{\bf r}_n }
		\cos ({\bf Q}\cdot{\bf r}_n) 
- \notag{}\\
&&-
\frac{({\bf u}\cdot {\bf k} )^2}{4}
\sum_{ {\bf r}_n } 
	e^{ - i {\bf k}\cdot{\bf r}_n } 
	\left[1 + \cos (2{\bf Q}\cdot{\bf r}_n)\right] + \dots
\end{eqnarray}
The terms in
Eq.~(\ref{eq::rho_k})
can be combined as follows:
\begin{eqnarray}
%%%%%%%%%%%%%%%%%%%%%%%%%%%%%%%%%%%%%%%%%%%%%%%%%%
\label{eq::rho_mod}
%%%%%%%%%%%%%%%%%%%%%%%%%%%%%%%%%%%%%%%%%%%%%%%%%% 
\rho_{\bf k} 
=
\left[1 - \frac{({\bf u}\cdot {\bf k} )^2}{4} \right]
\rho_{0 {\bf k}}
+
\rho^{\bf Q}_{\bf k} 
+
\rho^{2 \bf Q}_{\bf k} + \dots
\end{eqnarray}
where
\begin{eqnarray}
%%%%%%%%%%%%%%%%%%%%%%%%%%%%%%%%%%%%%%%%%%%%%%%%%%
\label{eq::cdw_peak1}
%%%%%%%%%%%%%%%%%%%%%%%%%%%%%%%%%%%%%%%%%%%%%%%%%% 
\rho^{\bf Q}_{\bf k} 
&=&
- i ({\bf u}\cdot {\bf k} ) 
\sum_{ {\bf r}_n } 
	e^{ - i {\bf k}\cdot{\bf r}_n } \cos ({\bf Q}\cdot{\bf r}_n)
=
\notag{} \\
&=&- \frac{i}{2} ({\bf u}\cdot {\bf k} )
F_{\bf k} \sum_{ {\bf b} } 
	(\delta_{{\bf k}, {\bf b+Q}} 
	+ 
	\delta_{{\bf k}, {\bf b-Q}}),
\\
%%%%%%%%%%%%%%%%%%%%%%%%%%%%%%%%%%%%%%%%%%%%%%%%%%
\label{eq::cdw_peak2}
%%%%%%%%%%%%%%%%%%%%%%%%%%%%%%%%%%%%%%%%%%%%%%%%%% 
\rho^{2 \bf Q}_{\bf k} 
&=&
-
\frac{({\bf u}\cdot {\bf k} )^2}{4}
\sum_{ {\bf r}_n } 
	e^{ - i {\bf k}\cdot{\bf r}_n } 
	\cos (2 {\bf Q}\cdot{\bf r}_n) =
\notag{}
\\
	&=&
-
\frac{({\bf u}\cdot {\bf k} )^2}{8}
F_{\bf k}  \sum_{ {\bf b} } 
	(\delta_{{\bf k}, {{\bf b}+2{\bf Q}}} 
	+ 
	\delta_{{\bf k}, {{\bf b}-2{\bf Q}}}).
\end{eqnarray} 
These terms describe appearance of the CDW peaks with wave-vectors
$n{\bf Q}, n = 1,2,\dots$

Of particular interest to us is the first term in
Eq.~(\ref{eq::rho_mod}).
We note that the presence of the CDW suppresses the amplitudes of the
Bragg peaks by an amount
\begin{eqnarray}
%%%%%%%%%%%%%%%%%%%%%%%%%%%%%%%%%%%%%%%%%%%%%%%%%%
\label{eq::delta_rho}
%%%%%%%%%%%%%%%%%%%%%%%%%%%%%%%%%%%%%%%%%%%%%%%%%%
\delta \rho^{\rm Bragg}_{\bf k}
=
- \frac{({\bf u}\cdot {\bf k} )^2}{4} \rho_{0 {\bf k}}.
\end{eqnarray}
We use this relation in
Sec.~\ref{subsect::Bragg_peak}.

It is important for our analysis that
Eq.~(\ref{eq::delta_rho})
remains valid also when CDW correlations are only short-ranged, while the
true long-range CDW order is absent. To show this, we consider more general
expression for the ionic density:
\begin{eqnarray}
%%%%%%%%%%%%%%%%%%%%%%%%%%%%%%%%%%%%%%%%%%%%%%%%%%
\label{eq::cdw_with_phonons}
%%%%%%%%%%%%%%%%%%%%%%%%%%%%%%%%%%%%%%%%%%%%%%%%%% 
\rho ( {\bf r} ) 
=
\sum_{ {\bf r}_n } 
	\delta \left[ {\bf r} - {\bf r}_n - 
		{\bf u} \cos ({\bf Q}\cdot{\bf r}_n + \phi ({\bf r}_n))
\right].
\end{eqnarray}
Here
$\phi ({\bf r}_n)$
is the phase of the order parameter. We assume that
$\phi$ is a slowly
varying function of 
${\bf r}_n$.
These variations are often referred to as ``phasons''. When 
$\phi ({\bf r}_n)$
varies as a function of 
${\bf r}_n$,
the CDW order weakens, or disappears completely, and becomes replaced by
short-range correlations. Generalizing 
Eq.~(\ref{eq::rho_k})
to account for the hot phonons, we derive
\begin{eqnarray}
%%%%%%%%%%%%%%%%%%%%%%%%%%%%%%%%%%%%%%%%%%%%%%%%%% 
\label{eq::rho_k_cdw_phon}
%%%%%%%%%%%%%%%%%%%%%%%%%%%%%%%%%%%%%%%%%%%%%%%%%% 
\rho_{\bf k} &&=
\left[ 1 - \frac{({\bf u}\cdot {\bf k} )^2}{4} \right]
\rho_{0 {\bf k}} 
\\
\nonumber
&&-
i ({\bf u}\cdot {\bf k} ) 
	\sum_{ {\bf r}_n } 
		e^{ - i {\bf k}\cdot{\bf r}_n }
		\cos ({\bf Q}\cdot{\bf r}_n + \phi ({\bf r}_n)) 
\\
\nonumber 
&&-
\frac{({\bf u}\cdot {\bf k} )^2}{4}
\sum_{ {\bf r}_n } 
	e^{ - i {\bf k}\cdot{\bf r}_n } 
	\cos (2{\bf Q}\cdot{\bf r}_n + 2 \phi ({\bf r}_n)).
\end{eqnarray}
Thus, in the presence of the phase variation
$\phi ({\bf r}_n)$
the amplitudes of the Bragg peaks remain unchanged, c.f.
Eq.~(\ref{eq::rho_mod}),
while the CDW amplitudes
$\rho^{\bf Q}_{\bf k}$
and
$\rho^{2\bf Q}_{\bf k}$
become equal to
\begin{flalign}
%%%%%%%%%%%%%%%%%%%%%%%%%%%%%%%%%%%%%%%%%%%%%%%%%% 
\label{eq::rho_phi}
%%%%%%%%%%%%%%%%%%%%%%%%%%%%%%%%%%%%%%%%%%%%%%%%%% 
\rho^{\bf Q}_{\bf k} 
=
- i ({\bf u}\cdot {\bf k} ) \sum_{ {\bf r}_n } 
		e^{ - i {\bf k}\cdot{\bf r}_n }
		\cos ({\bf Q}\cdot{\bf r}_n + \phi ({\bf r}_n)),
\\
%%%%%%%%%%%%%%%%%%%%%%%%%%%%%%%%%%%%%%%%%%%%%%%%%% 
\label{eqn::App_CDW_peak}
%%%%%%%%%%%%%%%%%%%%%%%%%%%%%%%%%%%%%%%%%%%%%%%%%% 
\rho^{2 \bf Q}_{\bf k} 
=
-
\frac{({\bf u}\cdot {\bf k} )^2}{4}
\sum_{ {\bf r}_n } 
	e^{ - i {\bf k}\cdot{\bf r}_n } 
	\cos (2 {\bf Q}\cdot{\bf r}_n + 2\phi ({\bf r}_n)).
\end{flalign} 
We, therefore, conclude that the CDW-induced changes in the Bragg peaks
intensities carry information about the short-range CDW correlations. In
particular, by means of
Eq.~(\ref{eq::delta_rho})
one can extract the amplitude
$u$ experimentally.

\section{Small oscillations near equilibrium state}
%%%%%%%%%%%%%%%%%%%%%%%%%%%%%%%%%%%%%%%%%%%%%%%%%%
\label{app::small_oscillations}
%%%%%%%%%%%%%%%%%%%%%%%%%%%%%%%%%%%%%%%%%%%%%%%%%% 

The TDGL sector of our formalism,
Eqs.~(\ref{eq::TDGL1})
and~(\ref{eq::TDGL2}),
contains several unknown coefficients. An important part of our study is
the evaluation of these parameters consistent with the available data. An
interesting possibility in this regard is to investigate the regime of
small oscillations of
$x$ and
$y$ near the equilibrium state. The resulting
theoretically determined frequency and damping factor can be compared with
experimental data for the AM oscillation spectrum, which allows us to
recover several parameters of our model. Since the calculations for
$T < T_{\rm c}$
and
$T > T_{\rm c}$
differ, they will be presented separately.

\subsection{Oscillations for
$T < T_{\rm c}$}

When
$T < T_{\rm c}$,
both order parameters
$x$ and
$y$ have non-zero values at equilibrium. In
this regime, we parameterize small oscillations as
\begin{eqnarray}
x = \sqrt{{\Theta}} + \delta x,
\quad
y = \sqrt{{\Theta}} + \delta y,
\end{eqnarray} 
where both 
$\delta x$
and 
$\delta y$
are complex variables. Writing
$\delta x$
and 
$\delta y$
as sums of real and imaginary parts
$\delta x = \delta x' + i \delta x''$
and
$\delta y = \delta y' + i \delta y''$,
we derive the following system of linearized equations
\begin{eqnarray}
%%%%%%%%%%%%%%%%%%%%%%%%%%%%%%%%%%%%%%%%%%%%%%%%%%
\label{eq::real_1}
%%%%%%%%%%%%%%%%%%%%%%%%%%%%%%%%%%%%%%%%%%%%%%%%%% 
&&\tau_0 \frac{d \delta x'}{dt} + 2{\Theta} \delta x'
				+ \zeta (\delta x' - \delta y') = 0,
\\
&&\frac{1}{\omega_0^2} \frac{d^2 \delta y'}{dt^2} 
	+ \frac{\gamma_y}{\omega_0} \frac{d \delta y'}{dt} 
	+ (\delta y' - \delta x') = 0,
%%%%%%%%%%%%%%%%%%%%%%%%%%%%%%%%%%%%%%%%%%%%%%%%%%
\label{eq::real_2}
%%%%%%%%%%%%%%%%%%%%%%%%%%%%%%%%%%%%%%%%%%%%%%%%%% 
\\
&&\tau_0 \frac{d \delta x''}{dt} + \zeta (\delta x'' - \delta y'') = 0,
%%%%%%%%%%%%%%%%%%%%%%%%%%%%%%%%%%%%%%%%%%%%%%%%%%
\label{eq::imag_1}
%%%%%%%%%%%%%%%%%%%%%%%%%%%%%%%%%%%%%%%%%%%%%%%%%% 
\\
&&\frac{1}{\omega_0^2} \frac{d^2 \delta y'' }{dt^2} 
	+ \frac{\gamma_y}{\omega_0} \frac{d \delta y''}{dt} 
	+ (\delta y'' - \delta x'') = 0.
%%%%%%%%%%%%%%%%%%%%%%%%%%%%%%%%%%%%%%%%%%%%%%%%%%
\label{eq::imag_2}
%%%%%%%%%%%%%%%%%%%%%%%%%%%%%%%%%%%%%%%%%%%%%%%%%% 
\end{eqnarray}
In these equations, the dynamics of the real and imaginary components are
decoupled from each other.

\begin{figure}
\begin{minipage}[h]{1\linewidth}\ \ \ \ \ \ \ \ \ \ \ \ \ \ \ \ \ \ \ \ \ \ \ \ \ \ \ \ \ \ \ \ \ \ \ \ 
\ \ \ 
\center{\includegraphics[width=1\linewidth]{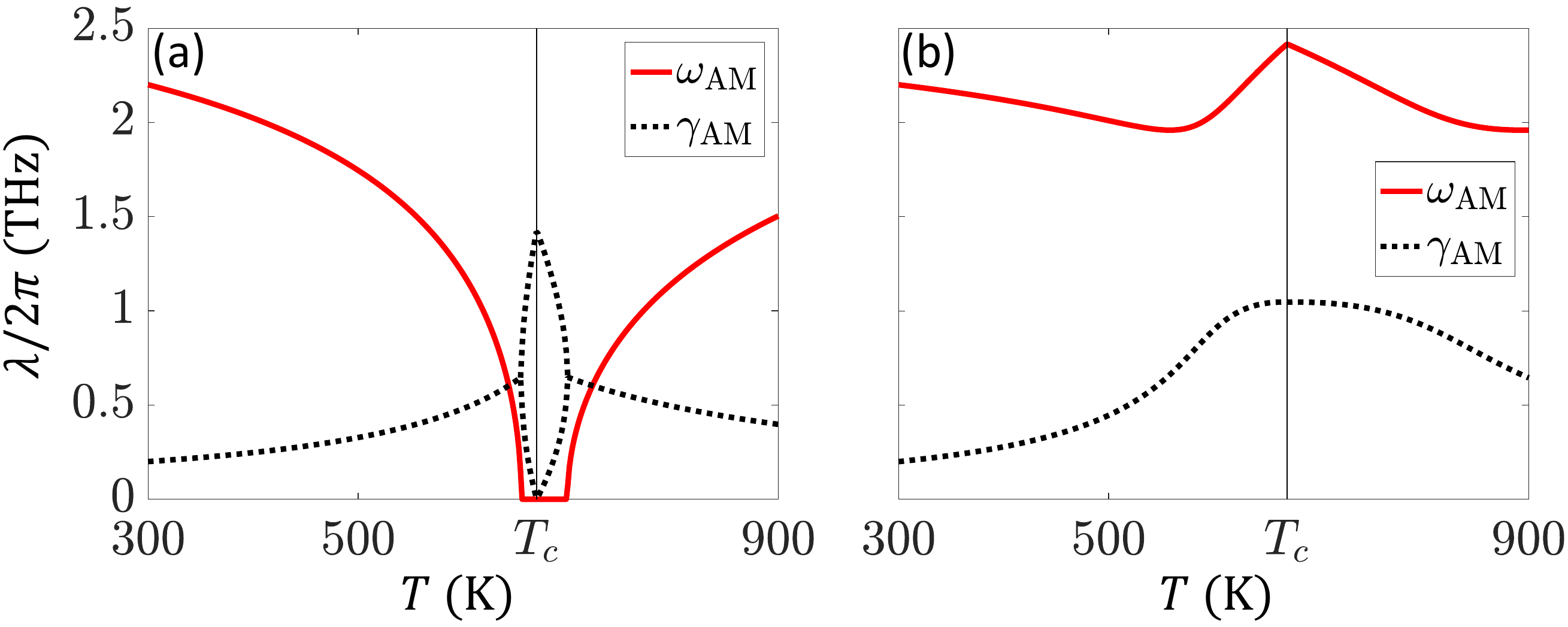}} 
\end{minipage}
\caption{Temperature dependence of the amplitude mode frequency
$\omega_{\rm AM}$
and damping parameter
$\gamma_{\rm AM}$
obtained by solving
Eqs.~(\ref{eq::freq}),~(\ref{eq::P_def}).
(a)~Adiabatic regime: the parameters in
Eq.~(\ref{eq::P_def})
are
$\omega_0/(2\pi) = 3.1$\,THz,
$\tau_0 = 20$\,fs,
$\zeta \approx 1.1$,
and
$\gamma_y \approx 0.04$.
These parameter values were used in our simulations, see
Table~\ref{tab::params_vals}.
They correspond to point~(a) marked in
Fig.~\ref{fig::Phase_diag}.
In a small temperature range around 
$T_{\rm c}$,
the amplitude oscillation mode turns into two overdamped modes with 
$\omega_{\rm AM}=0$
and unequal values of
$\gamma_{\rm AM}$
represented by two split dashed lines.
(b)~Non-adiabatic regime: the parameters in
Eq.~(\ref{eq::P_def}) 
are
$\omega_0/(2\pi) = 2.6$\,THz,
$\tau_0 = 40$\,fs,
$\zeta \approx 0.5$,
and
$\gamma_y \approx 0.033$.
This choice of parameters corresponds to point~(b) marked in
Fig.~\ref{fig::Phase_diag}.
%%%%%%%%%%%%%%%%%%%%%%%%%%%%%%%%%%%%%%%%%%%%%%%%%% 
\label{fig::frequencies}
%%%%%%%%%%%%%%%%%%%%%%%%%%%%%%%%%%%%%%%%%%%%%%%%%%
}
\end{figure}

We analyze first the frequencies of the real components in
Eqs.~(\ref{eq::real_1})
and~(\ref{eq::real_2}).
The insertion of an ansatz
$\delta x' (t) = X e^{ \lambda t }$
and
$\delta y' (t) = Y e^{ \lambda t }$
leads to an equation for
$\lambda$:
\begin{eqnarray}
%%%%%%%%%%%%%%%%%%%%%%%%%%%%%%%%%%%%%%%%%%%%%%%%%%
\label{eq::freq}
%%%%%%%%%%%%%%%%%%%%%%%%%%%%%%%%%%%%%%%%%%%%%%%%%% 
P( \lambda ) = 0,
\end{eqnarray}
where 
$P (\lambda )$
is a cubic polynomial defined as
\begin{eqnarray} 
%%%%%%%%%%%%%%%%%%%%%%%%%%%%%%%%%%%%%%%%%%%%%%%%%% 
\label{eq::P_def}
%%%%%%%%%%%%%%%%%%%%%%%%%%%%%%%%%%%%%%%%%%%%%%%%%%
P (\lambda ) = \tau_0 \lambda^3 +
(2 {\Theta} + \zeta + \gamma_y\omega_0 \tau_0) \lambda^2
+
\\
\nonumber 
\omega_0(2 {\Theta} \gamma_y + \zeta \gamma_y + \omega_0 \tau_0) \lambda 
+ 2 {\Theta} \omega_0^2.
\end{eqnarray}
Among three roots of 
$P ( \lambda )$,
one is always real and negative. Dependent on parameters, two other roots
are either (i)~both complex and conjugated to each other, or (ii)~both real
negative. In case~(i), the pair of complex roots represents AM. We identify
${\rm Im} \lambda$
with the frequency
$\omega_{\rm AM}$,
while
$-{\rm Re} \lambda$
is the AM damping parameter
$\gamma_{\rm AM}$.
The calculated values of
$\omega_{\rm AM}$
and
$\gamma_{\rm AM}$,
as functions of temperature, are plotted in
Fig.~\ref{fig::frequencies}.

Examining panels~(a) and~(b) of
Fig.~\ref{fig::frequencies}
we notice that the temperature dependence of the AM exhibits two different
qualitative regimes determined by the model parameters. Frequency
$\omega_{\rm AM}$
plotted in panel~(a) remains zero in some finite vicinity of 
$T_{\rm c}$.
As for panel~(b), 
$\omega_{\rm AM}$
never vanishes. Following
Refs.~\citenum{Schafer_PRL_2010,Schaefer_PRB_2014},
where this dichotomy was previously analyzed, we refer to the behavior
shown in panel~(a) of
Fig.~\ref{fig::frequencies}
as ``adiabatic'', while the one shown is in panel~(b) is to be called
``non-adiabatic''.

To determine the border between the adiabatic and non-adiabatic regimes, we
need to analyze
$\omega_{\rm AM}$
at
$T=T_{\rm c}$.
This condition corresponds to 
$\Theta = 0$.
As a result,
Eq.~(\ref{eq::freq})
becomes easily solvable:
\begin{eqnarray}
%%%%%%%%%%%%%%%%%%%%%%%%%%%%%%%%%%%%%%%%%%%%%%%%%%
\label{eq::tc_roots}
%%%%%%%%%%%%%%%%%%%%%%%%%%%%%%%%%%%%%%%%%%%%%%%%%% 
\lambda_1 = 0,
\quad
\lambda_{2,3} = 
-\frac{1}{2\tau_0}
\left[ 
	\gamma_y \omega_0 \tau_0 + \zeta
	\pm \sqrt{ {\cal D} }
\right], 
\end{eqnarray}
where
\begin{eqnarray} 
%%%%%%%%%%%%%%%%%%%%%%%%%%%%%%%%%%%%%%%%%%%%%%%%%%
\label{eq::D_def}
%%%%%%%%%%%%%%%%%%%%%%%%%%%%%%%%%%%%%%%%%%%%%%%%%% 
{\cal D} = (\gamma_y \omega_0 \tau_0 - \zeta )^2 - 4 (\omega_0\tau_0)^2.
\end{eqnarray} 
Quantity 
${\cal D}$
is important for our analysis. Specifically,
Eq.~(\ref{eq::tc_roots})
implies that, if
${\cal D} > 0$,
then
$\omega_{\rm AM} = 0$,
otherwise, it is finite:
$\omega_{\rm AM}
= 
| {\rm Im} \lambda_{2,3} | = \frac{1}{2\tau_0} \sqrt{-\cal D}$.
Therefore, the condition
${\cal D} = 0$
separates the adiabatic and non-adiabatic regimes. (In terms of experiment,
it might be difficult to detect the difference between a formally adiabatic
case
$\omega_{\rm AM} (T=T_{\rm c}) = 0$,
and a non-adiabatic case characterized by inequality
$\gamma_{\rm AM} (T=T_{\rm c}) \gg \omega_{\rm AM} (T=T_{\rm c})$.)

Since the values of
$\omega_{\rm AM}$
and
$\gamma_{\rm AM}$
at
$T=300$\,K
are known from experiment, see
Eq.~(\ref{eq::AM_param}),
we can use them to derive constraints on the TDGL parameters. To obtain the
constraints, we re-write
Eq.~(\ref{eq::freq}) 
as two real-valued equations
\begin{eqnarray}
%%%%%%%%%%%%%%%%%%%%%%%%%%%%%%%%%%%%%%%%%%%%%%%%%%
\label{eq::TDGL_params_constraints1}
%%%%%%%%%%%%%%%%%%%%%%%%%%%%%%%%%%%%%%%%%%%%%%%%%% 
{\rm Re}\, P (i \omega_{\rm AM} - \gamma_{\rm AM})
= 0,
\\
%%%%%%%%%%%%%%%%%%%%%%%%%%%%%%%%%%%%%%%%%%%%%%%%%%
\label{eq::TDGL_params_constraints2}
%%%%%%%%%%%%%%%%%%%%%%%%%%%%%%%%%%%%%%%%%%%%%%%%%% 
{\rm Im}\, P (i \omega_{\rm AM} - \gamma_{\rm AM})
= 0.
\end{eqnarray} 
These equations reduce the number of free TDGL coefficients from four
($\omega_0$,
$\tau_0$,
$\zeta$,
and
$\gamma_y$)
to two. In the main text, we treat
$\tau_0$
and
$\omega_0$
as free parameters. Within such a convention, 
Eqs.~(\ref{eq::TDGL_params_constraints1})
and~(\ref{eq::TDGL_params_constraints2})
can be used to define two implicit functions
$\gamma_y = \gamma_y (\omega_0, \tau_0)$
and
$\zeta = \zeta (\omega_0, \tau_0)$.

Damping parameter
$\gamma_y$
must always be non-negative, i.e.
\begin{eqnarray}
\gamma_y (\omega_0, \tau_0) \geq 0,
\end{eqnarray} 
which further limits the available space for
$\omega_0$
and
$\tau_0$
as discussed in the main text, see also
Fig.~\ref{fig::Phase_diag}.

Now we can analyze
Eqs.~(\ref{eq::imag_1})--(\ref{eq::imag_2}),
which describe the oscillations of the imaginary components 
$\delta x''$
and
$\delta y''$.
One can check that, in this case, we also have three modes whose
eigenfrequencies are given by
Eq.~(\ref{eq::tc_roots}).
The zero eigenfrequency represents a Goldstone mode. Within our model small
oscillations of
$\delta x''$
and
$\delta y''$
have temperature-independent frequencies and damping parameters. This is a
consequence of our assumption that quantity
$a$ in the Landau
functional~(\ref{eq::landau_func})
is the only one dependent on temperature.

As with the real components, the dynamics of the imaginary components
$\delta x''$
and
$\delta y''$
is sensitive to the sign of
${\cal D}$.
Specifically, in the adiabatic regime, the roots
$\lambda_{2,3}$
are both real negative, i.e. the time evolution is overdamped. In the
non-adiabatic regime, the roots form a complex conjugate pair, which
corresponds to underdamped oscillations.

\subsection{Oscillations for
$T > T_{\rm c}$}

When
$T \geq T_{\rm c}$,
the equilibrium values of
$x$ and
$y$ are zero. Thus, in the regime of
linear oscillations in the disordered phase one writes
$x(t) = \delta x(t)$
and
$y(t) = \delta y(t)$.
The resulting linearized equations for the real components coincide with
the equations for the imaginary components:
\begin{eqnarray}
&&\tau_0 \frac{d \delta x }{dt} -{\Theta} \delta x
				+ \zeta (\delta x - \delta y) = 0,
\\
&&\frac{1}{\omega_0^2} \frac{d^2 \delta y}{dt^2} 
	+ \frac{\gamma_y}{\omega_0} \frac{d \delta y}{dt} 
	+ (\delta y - \delta x) = 0.
\end{eqnarray}
The eigenfrequencies then satisfy the equation
\begin{eqnarray}
%%%%%%%%%%%%%%%%%%%%%%%%%%%%%%%%%%%%%%%%%%%%%%%%%%
\label{eq::small_osc_im}
%%%%%%%%%%%%%%%%%%%%%%%%%%%%%%%%%%%%%%%%%%%%%%%%%% 
\tau_0 \lambda^3 &+& (\zeta - {\Theta} + \gamma_y\omega_0 \tau_0) \lambda^2
\\
\nonumber 
&+& \omega_0 (\gamma_y \zeta - \gamma_y {\Theta} + \omega_0 \tau_0) \lambda
- {\Theta} \omega_0^2 =0.
\end{eqnarray}
Naturally, at the transition
(${\Theta} = 0$),
Eq.~(\ref{eq::small_osc_im}) 
and
Eq.~(\ref{eq::freq})
are identical. This ensures that all eigenfrequencies smoothly cross 
$T_{\rm c}$.
Figure~\ref{fig::frequencies}
shows the numerically calculated eigenfrequencies for both adiabatic and
non-adiabatic regimes.

\section{Electronic heat capacity}
%%%%%%%%%%%%%%%%%%%%%%%%%%%%%%%%%%%%%%%%%%%%%%%%%% 
\label{app::C_el}
%%%%%%%%%%%%%%%%%%%%%%%%%%%%%%%%%%%%%%%%%%%%%%%%%% 

\subsection{Temperature dependence of the electronic heat capacity}
%%%%%%%%%%%%%%%%%%%%%%%%%%%%%%%%%%%%%%%%%%%%%%%%%%
\label{app_subsect::pseudogap_heat_cap}
%%%%%%%%%%%%%%%%%%%%%%%%%%%%%%%%%%%%%%%%%%%%%%%%%% 
Here we further motivate 
Eq.~(\ref{eq::electr_heat_cap})
for the temperature dependence of the electronic heat capacity. We already
mentioned that
LaTe$_3$,
despite the presence of the CDW order, is not an insulator, but rather is a
metal, with ungapped fragments of the Fermi surface and finite density of
states at the Fermi energy
$\epsilon_F = 0$,
cf.
Fig.~\ref{fig:ARPES}.
Consequently, the low-temperature heat capacity
demonstrates~[\!\!~\citenum{ru_tritell_transp_theromod2006}]
metallic behavior expressed by
Eq.~(\ref{eq::electr_heat_cap0}).
However, unlike a ``classical" metal for which deviations from linear
relation
$C_{\rm e} \propto T_{\rm e}$
for
$T_{\rm e} \lesssim 2000$\,K
are generally
weak~[\!\!~\citenum{lin_2temp_model2008}],
we expect that the CDW order in
LaTe$_3$
affects the validity of
Eq.~(\ref{eq::electr_heat_cap0})
in the above temperature range.
\begin{figure}[t!]
\includegraphics[width=9cm]{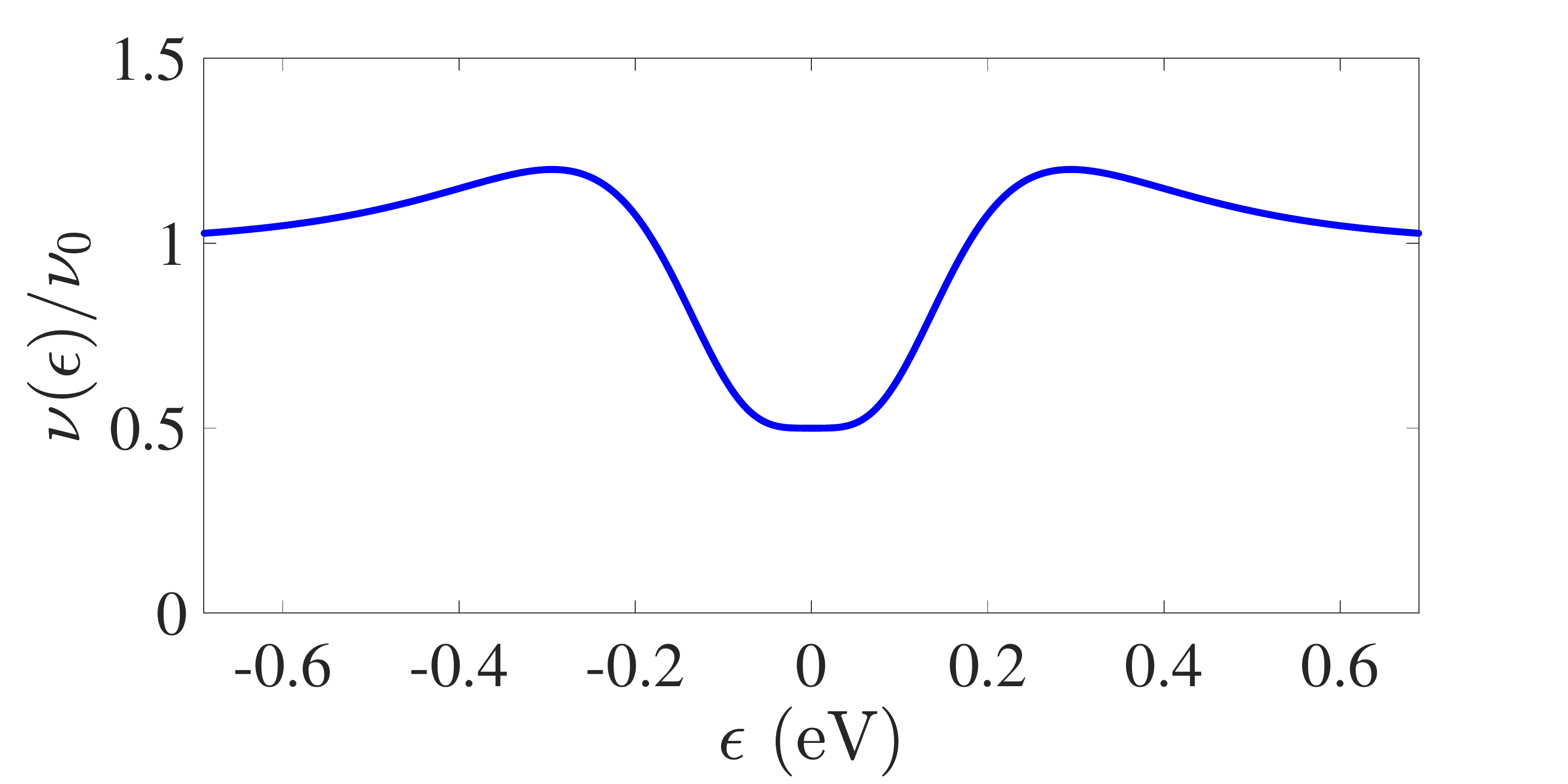}
\caption{Sketch of DOS in the CDW state. It represents the vicinity 
of a Fermi surface that has both metallic and gaped regions. The 
gapped regions are responsible  for the suppression of the DOS at
$\epsilon_F = 0$
and the appearance of the maxima around 
$|\varepsilon| \sim \Delta \sim 350$\,meV.
%%%%%%%%%%%%%%%%%%%%%%%%%%%%%%%%%%%%%%%%%%%%%%%%%% 
\label{fig::dos_modif}
%%%%%%%%%%%%%%%%%%%%%%%%%%%%%%%%%%%%%%%%%%%%%%%%%% 
}
\end{figure}
\begin{figure}[t!]
\includegraphics[width=9cm]{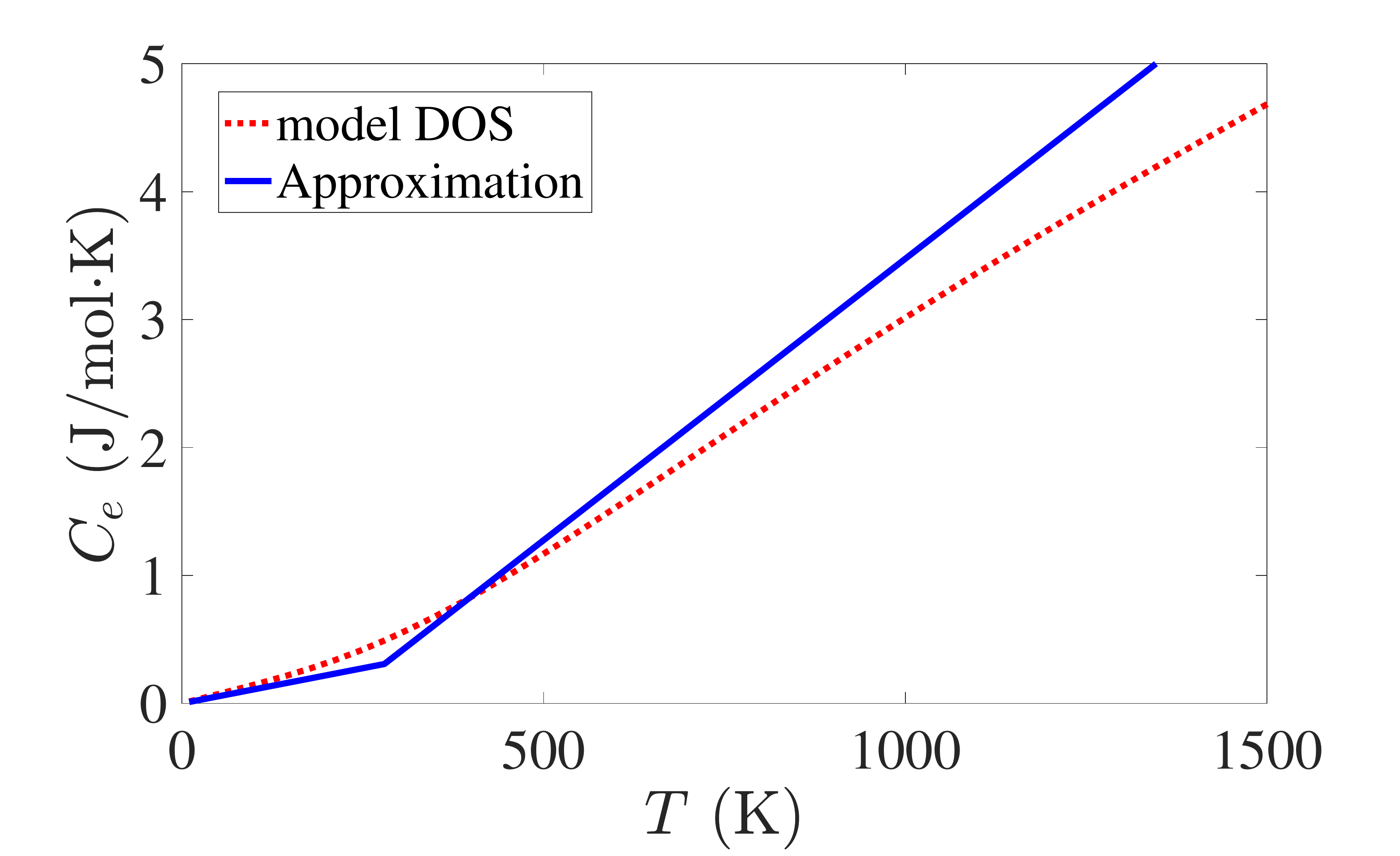}
\caption{Dashed (red) line represents temperature dependence of the
electronic heat capacity for the DOS shown in
Fig.~\ref{fig::dos_modif}.
Solid (blue) line corresponds to our approximation
Eq.~(\ref{eq::electr_heat_cap0}).
%%%%%%%%%%%%%%%%%%%%%%%%%%%%%%%%%%%%%%%%%%%%%%%%%% 
\label{fig::cv}
%%%%%%%%%%%%%%%%%%%%%%%%%%%%%%%%%%%%%%%%%%%%%%%%%% 
}
\end{figure}
\begin{figure}[h]
\includegraphics[width=1\linewidth]{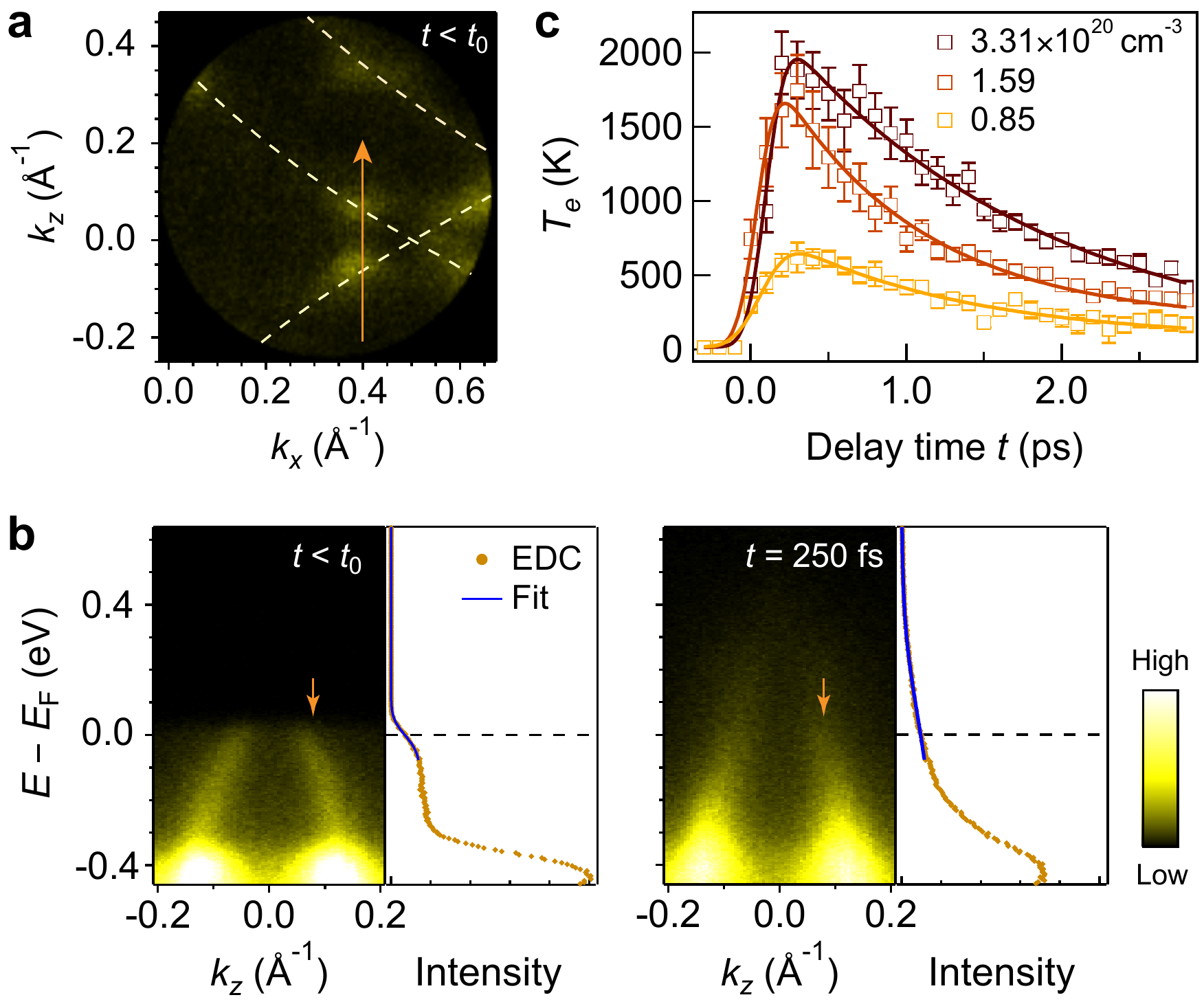}
\caption{Estimating electronic temperature
$T_{\rm e}$
after a photoexcitation. (a)~Fermi surface map before arrival of a laser
pulse
($t<t_0$).
Intensities are integrated over
$\pm10$\,meV
around the Fermi energy
$\epsilon_F$.
Dashed curves represent calculated Fermi surface  based on a tight-binding
model~[\!\!~\citenum{PhysRevLett.93.126405}],
in the absence of the CDW order. The arrow marks the energy-momentum cut
through the ungapped part of the Fermi surface displayed in~(b). In~(b),
the band dispersions are shown at two representative pump-probe time
delays: before the laser pulse (left) and
250\,fs
after the pulse arrival (right). The energy-distribution curves (EDCs) are
obtained by integrating over a window
$\Delta k = 0.05$\,\AA$^{-1}$
at the momentum indicated by the arrow. Blue curve is a fit by 
Eq.~\eqref{eq:Te_ARPES}
to a part of the EDC. The Fermi energy is indicated by the dashed line.
Data in (a) and (b) were obtained at a photoexcitation density of
$3.31\times10^{20}$\,cm$^{-3}$.
(c)~Electronic temperature
$T_{\rm e}$
plotted as a function of the pump-probe delay for the three photoexcitation
densities indicated in the plot legend.  Curves are the fits to a
single-exponential decay
model~[\!\!~\citenum{Alfred2018}].
Error bars represent one standard deviation of the fits.
%%%%%%%%%%%%%%%%%%%%%%%%%%%%%%%%%%%%%%%%%%%%%%%%%% 
\label{fig:ARPES}
%%%%%%%%%%%%%%%%%%%%%%%%%%%%%%%%%%%%%%%%%%%%%%%%%% 
}
\end{figure}
Available thermodynamic and {\it ab initio} data 
support~[\!\!~\citenum{ru_tritell_transp_theromod2006}]
this expectation: the CDW order suppresses coefficient 
$c_0$
in
Eq.~(\ref{eq::electr_heat_cap0})
almost twofold relative to its value in the hypothetical situation without
the CDW order. We assume that this suppression is due to the
``pseudo-gapped" single-electron density of state (DOS) 
$\nu (\epsilon)$
sketched in
Fig.~\ref{fig::dos_modif}.
In this sketch, the uniform metallic DOS
$\nu_0$
is modified by the presence of the CDW order. At the Fermi energy
$\epsilon_F$,
we choose it for concreteness to be two times smaller than the bare value
$\nu_0$.
This suppression is caused by the expulsion of the electronic states from
the gapped parts of the Fermi surface to higher energies. Since coefficient
$c_0 \propto \nu (\epsilon_F)$,
the value of
$c_0$
decreases together with 
$\nu (\epsilon_F)$.
As for the states excluded from the vicinity of
$\epsilon_F$,
they accumulate at~[\!\!~\citenum{hu_3Te_optics2014}]
$|\epsilon - \epsilon_F| \sim \Delta \sim 0.35$\,eV.
When
$|\epsilon - \epsilon_F| \gtrsim 0.6$\,eV,
the DOS returns to its bare value 
$\nu_0$.
The resulting function
$\nu ( \epsilon )$
exhibits pronounced variations on the scale of hundreds of meV, which leads
to a non-linear temperature-dependence of the heat capacity shown in
Fig.~\ref{fig::cv}.
This temperature dependence is calculated using the expression:
\begin{eqnarray}
%%%%%%%%%%%%%%%%%%%%%%%%%%%%%%%%%%%%%%%%%%%%%%%%%%
\label{eq::cv}
%%%%%%%%%%%%%%%%%%%%%%%%%%%%%%%%%%%%%%%%%%%%%%%%%% 
C_{\rm e} (T_{\rm e})
=
 \int_{-\infty}^{+\infty}
 	\frac{ \nu (\varepsilon) \varepsilon^2 d \varepsilon }
	{2 T_{\rm e}^2 \cosh^2( \varepsilon /(2T_{\rm e}))}.
\end{eqnarray} 
The plot in
Fig.~\ref{fig::cv}
indicates that 
$C_{\rm e} (T_{\rm e})$
departs from the low-temperature linear dependence,
Eq.~(\ref{eq::electr_heat_cap0}),
for
$T \gtrsim 300$\,K.
Figure~\ref{fig::cv}
also shows the plot of
$C_{\rm e} (T_{\rm e})$
for a simple piecewise linear function given by
Eq.~(\ref{eq::electr_heat_cap}).
It can be seen in this figure that
Eq.~(\ref{eq::electr_heat_cap})
adequately approximates
$C_{\rm e} (T_{\rm e})$
in the temperature range of interest.

We note that
expression~(\ref{eq::cv})
is formulated under the assumption that
$\nu ( \varepsilon )$
is independent of 
$T_{\rm e}$.
This assumption is, likely, violated in
LaTe$_3$,
because the actual DOS is sensitive to the values of the order parameters
$x$ and
$y$, both of which are temperature- and time-dependent quantities.
Thus, application of
Eq.~(\ref{eq::electr_heat_cap})
to the non-equilibrium situations should be taken with caution. However, we
do expect that the piecewise linear function
Eq.~(\ref{eq::electr_heat_cap})
would still constitute a reasonable approximation to the actual temperature
dependence of the electronic heat capacity. Let us also emphasize that
accurate knowledge of the electronic heat capacity is important only during
the first rapid stage of the electronic temperature relaxation. At the
second slow stage, only lattice contributions to the heat capacity are
relevant, see
Eq.~(\ref{eq::tau_DW}).

\subsection{Estimate of electronic temperature from tr-ARPES}
%%%%%%%%%%%%%%%%%%%%%%%%%%%%%%%%%%%%%%%%%%%%%%%%%%
\label{app_subsect::tr_ARPES}
%%%%%%%%%%%%%%%%%%%%%%%%%%%%%%%%%%%%%%%%%%%%%%%%%% 

The generation of hot carriers after strong photoexcitation is followed by
thermalization within the electronic subsystem on a
time scale
$\lesssim 100$\,fs~[\!\!~\citenum{Prasankumar2016}].
Using time- and angle-resolved photoemission spectroscopy (tr-ARPES), one
can estimate the electronic temperature after the initial thermalization by
fitting the energy distribution of quasiparticles to a Fermi-Dirac
distribution~[\!\!~\citenum{Wang2012}].

Figure~\ref{fig:ARPES}a
shows the Fermi surface in
LaTe$_3$
at
$T=15$\,K$\ll T_{\rm c}$
before the arrival of the pump laser pulse (see
Ref.~\citenum{Alfred2018}
for measurement details). The tr-ARPES intensity is absent for the most
parts of the Fermi surface due to the opening of the CDW gap. The remaining
Fermi surface is consistent with the previous
reports~[\!\!~\citenum{PhysRevLett.93.126405,Brouet2008}].
In order to minimize complications arising from the transient suppression
of the CDW gap, we focus on the ungapped part of the Fermi surface at
equilibrium. In
Fig.~\ref{fig:ARPES}b,
we present an energy-momentum cut through the metallic part of the Fermi
surface where the Te
$5p_x/p_z$
bands cross the Fermi level
$\epsilon_F$.
The same cut is shown after photoexcitation at a pump-probe time delay of
$t=250$\,fs,
where states above
$\epsilon_F$
are transiently populated. 

To quantitatively analyze the carrier redistribution after photoexcitation,
we plot the energy distribution curves (EDCs) at
$k_z=0.08$\,\AA$^{-1}$
indicated by the arrow in
Fig.~\ref{fig:ARPES}b.
At
$t<t_0$,
there is a sharp cutoff of EDC around
$\epsilon_F$;
this feature is replaced by a long tail at
$\epsilon>\epsilon_F$
at 250\,fs. The temporal evolution of the EDC across
$\epsilon_F$
can be captured by the following
model~[\!\!~\citenum{Fann1992,Wang2012}]:
\begin{equation}\label{eq:Te_ARPES}
I(\epsilon,t)
=
\{\nu(\epsilon)f[\epsilon,\mu(t),T_{\rm e}(t)]\}
\ast
\tilde{g}[\epsilon,\tilde{w}(t)],
\end{equation}
where
$\nu(\epsilon)$
is the density of states,
$f[\cdot]$
is the Fermi-Dirac distribution that depends on the chemical potential
$\mu$ and the electronic temperature
$T_{\rm e}$.
The terms in
$\{\cdot\}$
are energy convoluted with a Gaussian kernel
$\tilde{g}[\cdot]$,
cf.
Eq.~(\ref{eq::filter}), 
whose
time-dependent~[\!\!~\citenum{Ishida2011,Wang2012}]
width parameter
$\tilde{w}(t)$
arises from the finite energy resolution of the instrument and from
spectral broadening due to increased scattering rate after photoexcitation.
The density of states
$\nu(\epsilon)$
is assumed to remain unchanged over time; it is determined by the EDCs
before photoexcitation. This assumption is largely justified, because that
particular part of the Fermi surface is minimally affected by the transient
suppression and recovery of the CDW gap. To limit the number of free
parameters, we adopt a linear approximation
$\nu(\epsilon) \approx \nu (\epsilon_{\rm F})
+ (\epsilon - \epsilon_{\rm F}) \alpha_\nu$,
where
$\alpha_\nu$
is an adjustable parameter.
This, in turn, limits the fitting range as indicated in
Fig.~\ref{fig:ARPES}b,
since the strong intensity of a separate band at high binding energy cannot
be captured with a linear density of states. In summary, the time-dependent
fitting parameters include
$\mu(t)$,
$T_{\rm e}(t)$,
and
$\tilde{w}(t)$; 
the value of
$T_{\rm e}$
before the arrival of the laser pulse is fixed to be
15\,K,
which is the base temperature of the sample during the measurement.

Figure~\ref{fig:ARPES}c
shows the extracted
$T_{\rm e}$
throughout the photoexcitation event for three different excitation
densities given in terms of the number of absorbed photons per unit volume.
Though electronic temperature obtained within a few 
$\tau_0$
after photoexcitation is less reliable due to the non-thermal nature of the
carrier
distribution~[\!\!~\citenum{Prasankumar2016}],
values at longer time delays are indicative of the quasi-thermal state of
the electronic subsystem with effective electronic temperature 
$T_{\rm e}$
and a Fermi-Dirac distribution (see the fits in
Fig.~\ref{fig:ARPES}b).
As one expects, higher transient
$T_{\rm e}$
is reached at higher excitation density.

Now, using
Eq.~(\ref{eq::init_Te}) 
with the temperature dependence
$C_{\rm e} (T_{\rm e})$
given by
Eq.~(\ref{eq::electr_heat_cap})
and with the values of 
$\hbar \omega_\gamma$,
${\cal V}$,
$F$,
$T_{\rm env}$,
and
$T_{\rm e}$
corresponding to the experiment-based plots in
Fig.~\ref{fig:ARPES}c,
we obtain the possible range of values 
$3-5$\,mJ$/$mol\,K$^{2}$
for the parameter $c$ entering
Eq.~(\ref{eq::electr_heat_cap}).
In the actual simulations, we use
$c = 4$\,mJ$/$mol\,K$^{2}$.


\begin{thebibliography}{97}
\bibitem[{tim()}]{timeshift_note}
\bibinfo{note}{The limited temporal resolution in our experiments prevented us
  from accurately determining time zero -- the moment when the pump and the
  probe pulses overlap. The parameter $t_0$ is included in Eq.~(\ref{eq::BP})
  to account for the possible experimental inaccuracy of an overall shift in
  the delay time.}
\end{thebibliography}
\end{document}